\documentclass[aps,prb]{revtex4}
\usepackage{graphicx}
\begin{document}
\title{Dissipative particle dynamics for interacting systems}
\author{I. Pagonabarraga(*) and D. Frenkel}
\affiliation{FOM-Institute for Atomic and Molecular Physics (AMOLF),\\
Kruislaan 407, 1098 SJ Amsterdam (The Netherlands)\\
Current address: Departament de F\'{\i}sica Fonamental, Universitat de
Barcelona, Av. Diagonal 647, 08028-Barcelona, Spain}

\begin{abstract}
We introduce a dissipative particle dynamics scheme for the dynamics of
non-ideal fluids. Given a free-energy density that determines the
thermodynamics of the system, we derive consistent conservative forces. The
use of these effective, density dependent forces reduces the
local structure as compared to previously proposed models. This is an
important feature in mesoscopic modeling, since it ensures a realistic
length and time scale separation in coarse-grained models. We consider in
detail the behavior of a van der Waals fluid and a binary mixture with a
miscibility gap. We discuss the physical implications of having a single
length scale characterizing the interaction range, in particular for the
interfacial properties.
\end{abstract}

\maketitle

\section{Introduction}

There is a strong incentive to develop ``mesoscopic'' numerical techniques
to model the dynamics of fluids with different characteristic length scales.
Mesoscopic simulations make it possible to analyze processes that take place
on length and time scales that are out of reach for purely atomistic
simulations such as Molecular Dynamics (MD). In MD, one retains the full
atomic details in the description of the system, but at the expense of
restricting the studies to short times. In contrast, models that describe
the system at mesoscopic scales, employ a certain degree of coarse graining,
which allows one to analyze longer times. However, care should be taken that
the loss of ``atomic'' information associated with the coarse-graining
process does not lead to unrealistic features on larger length and time
scales. In particular, the coarse-grained models should provide an adequate
description of the equilibrium properties of the system. Some of the
mesoscopic models that have been proposed previously in the literature were
derived in a systematic way from underlying microscopic models, as is the
case with the lattice-Boltzmann method\cite{LB}, which can be viewed as a
preaveraged lattice gas model\cite{LG}. Coming from the opposite side,
smoothed particle dynamics was introduced as a Lagrangian discretization of
the macroscopic equations of fluid motion\cite{SPH}. A different strategy to
simulate structured fluids is to assume that the solvent is passive, and
that the suspended objects have a diffusive dynamics with diffusion
coefficients that are known {\em a priori}. This has led to the development
of Brownian\cite{EMc} and Stokesian dynamics\cite{BB}.

In the early nineties, Dissipative Particle Dynamics (DPD) was introduced as
a novel way to simulate fluids at a mesoscopic scale\cite{HK}. In DPD, the
fluid is represented by a large number ($N$) of point particles that have a
pairwise additive interaction. The interparticle forces are the sum of three
contributions. In addition to the usual conservative forces that can be
derived from a Hamiltonian, DPD includes dissipative and random forces.
These mimic the effect of viscous damping between fluid elements and the
thermal noise of the fluid elements, respectively. Flekk{\o}y and Coveney
\cite{Flekkoy} have shown that, in principle, a particular DPD-like model
can be derived from an atomistic description. However, no such derivations
exist for the commonly used DPD models. Nonetheless, even without such a
link to the underlying microscopics, it has been shown that thermal
equilibrium can be ensured by an appropriate choice of the ratio between
dissipative and random forces\cite{EW}. The hydrodynamic behavior of the DPD
model has been explored in some detail\cite{Colin,IF,Masters,Serrano},
although the link between the mesoscopic and the macroscopic description is
not completely understood.

In conventional DPD, all interparticle forces have the same finite
interaction range $r_{c}$. Their amplitudes decay according to a weight
function $w(r_{ij})$ that has been made to vanish at $r_{c}$ in order to
avoid spurious jumps at the cut-off distance. In this paper we employ a more
general description of the conservative interactions. In the existing
literature, the conservative forces have usually been assumed to depend
explicitly on the distance between a pair of particles. For the sake of
computational convenience, the conservative forces between DPD particles are
taken smooth and monotonic functions of the distance - in fact, the
smoothness of the forces is one of the advantages of DPD. When the forces
depend linearly on the interparticle separation, the equation of state (EOS)
of the DPD fluid is approximately quadratic in the density and exhibits no
fluid-fluid phase transition. Even though the forces between DPD particles
are smooth, they still induce structure in the fluid (reminiscent of atomic
behavior) on a length scale of order $r_{c}$. In this respect, the
conventional DPD scheme is similar to other mesoscopic models for non-ideal
fluids but differs from the - computationally more demanding - scheme of
Flekk{\o}y and Coveney that was mentioned above\cite{Flekkoy}.

Our aim in this paper is to arrive at a formulation of DPD that allows for a
description of the behavior of non-ideal fluids and fluid-mixtures.  To this
end, we look for a model in which there is a direct link between the
macroscopic equation of state and the effective interparticle forces. As we
will show, as an additional advantage,  our approach results in rather weak
structural correlations in the fluid. In the next section we describe in
detail the model and how conservative forces are derived. We will
subsequently elaborate the general method  on three characteristic examples:
a non-ideal fluid without a gas-liquid phase transition that has been
studied previously with a different choice of conservative forces, a van der
Waals fluid, and a mixture with a miscibility gap. In section \ref
{sect:inter} we look at the interfacial properties of these examples to gain
some insight in the physical meaning of the conservative forces that we
introduce, and subsequently analyze  their equilibrium behavior and compare
with previous models. We conclude with a discussion of our main results.

\section{Model}

\label{sect:model} In DPD one has $N$ point particles of mass $\{m_{i}\}$
that interact through a sum of pairwise-additive conservative, dissipative
and random forces. These particles can be interpreted as fluid elements, and
the dissipative forces are introduced to mimic the viscous drag between
them. The random force equilibrates the energy lost through friction between
the particles, enabling the system to reach an equilibrium state. To be
specific, if we call $\{{\bf r}_{k},{\bf p}_{k}\}$ the set of particle
positions and momenta of the $N$ point particles, their dynamics are
controlled by Newton equations of motion

\begin{eqnarray}
\frac{d{\bf r}_{k}}{dt} &=&{\bf v}_{k}  \label{eqts-DPD1} \\
\frac{d{\bf p}_{k}}{dt} &=&\sum_{j\neq i}\left\{ {\bf F}^{C}({\bf r}_{ij})+ 
{\bf F}^{D}({\bf r}_{ij})+{\bf F}^{R}({\bf r}_{ij})\right\}  \nonumber \\
&=&\sum_{j\neq i}\left\{ {\bf F}^{C}({\bf r}_{ij})-\gamma \omega ^{D}({\bf r}
_{ij}){\bf v}_{ij}\cdot {\bf e}_{ij}{\bf e}_{ij}+\sigma \omega ^{R}({\bf r}
_{ij}){\bf e}_{ij}\xi _{ij}\right\}  \label{eqts-DPD}
\end{eqnarray}
where we have used the notation ${\bf r}_{ij}\equiv {\bf r}_{i}-{\bf r}_{j}$
and ${\bf v}_{ij}\equiv {\bf v}_{i}-{\bf v}_{j}$. ${\bf e}_{ij}$ denotes a
unit vector in the direction of ${\bf r}_{ij}$, and ${\bf v}_{i}={\bf p}%
_{i}/m_{i}$ is the velocity of particle $i$. The dissipative force, ${\bf F}%
^{D}({\bf r}_{ij})$, depends both on the relative positions and velocities
of the interacting pair of particles and its amplitude is characterized by
the parameter $\gamma $. This parameter is related to the viscosity of the
DPD fluid. The third term in eq.({\ref{eqts-DPD}}), ${\bf F}^{R}( {\bf r}%
_{ij})$, is a random force acting on each pair of DPD particles - $\xi $
stands for a random variable with Gaussian distribution and unit variance.
The random force has an amplitude $\sigma $ and is also central. Central
pair interactions ensure angular momentum conservation (although the
dynamics can be generalized to account for non-central forces \cite{pepL}).
The dissipative and random forces are completely specified once the weight
functions, $\omega ^{D}(r_{ij})$ and $\omega ^{R}(r_{ij})$, are specified- these  are
smooth and of finite range. Although they can be chosen at will, Espa\~{n}ol
and Warren showed\cite{EW} that $\omega ^{D}$ and $\omega ^{R}$ must be
related to ensure that the probability to observe a particular configuration
of DPD particles is given by the Boltzmann distribution in equilibrium.
Specifically, if they are chosen such that $\omega ^{R}=%
\sqrt{\omega ^{D}}$, then the correct equilibrium distribution is recovered,
and the equilibrium temperature of the DPD fluid is fixed by the ratio of
the amplitudes of the dissipative and random forces, $k_{B}T=\sigma
^{2}/(2\gamma )$. We stress that the DPD equations of motion, eq.(\ref
{eqts-DPD1}-\ref{eqts-DPD}), cannot be derived from a Hamiltonian.

Traditionally, and for simplicity, the conservative forces in DPD have  been
taken as pairwise-additive and central, with a weight function related to $%
\omega ^{D}$, and with a variable amplitude that sets the temperature scale
in the system. As long as the force is sufficiently weak that it does not
induce appreciable inhomogeneities in the density around a DPD particle, it
can only lead to an equation of state with a quadratic dependence in the
density, irrespective of the precise choice for the weight function (see
below). One consequence is that phase separation between disordered phases
cannot occur in a pure system; at least a binary mixture of different kinds
of particles is needed\cite{phasesep}.

We will first consider the general form that the free energy of a DPD system
can have, in order to elucidate the generic shape of consistent conservative
forces. In agreement with the idea that the DPD particles refer to lumps of
fluid, it seems natural to assume that the relevant energy associated to
their configurations is a free energy, rather than a strictly ``mechanical''
potential energy. We can express quite generically the free energy ${\cal F}$
of an inhomogeneous system with density $\rho ({\bf r})$ as

\begin{equation}
{\cal F}=\int d{\bf r}\rho ({\bf r})f(n({\bf r}))  \label{freevol}
\end{equation}
where $f(\rho )$ is the expression for the local free energy per particle
(in units of $k_{B}T$), and $n(\{{\bf r}\})$ is related to the density of
the system at ${\bf r}$. This formulation is reminiscent of the strategy
followed in density functional theory to study the equilibrium properties of
the fluids \cite{Evans}. In fact, the particular case $n(\{{\bf r}\})=\rho
(\{{\bf r}\})$ corresponds to the local density approximation in density
functional theory, and if $n({\bf r})$ is chosen to be an average of the
density over an interval around ${\bf r}$, it can be understood as a
weighted density approximation for the true free energy. We
can separate the total free energy, $f(\rho)=f^{id}(\rho)+f^{ex}(\rho)$, as
the sum of the ideal $f^{id}(\rho )=\log (\Lambda ^{3}\rho)-1$ plus the
excess contribution, where $\Lambda $ is the thermal de-Broglie wavelength.
Our purpose is to obtain the equivalent expression for a DPD system, in
which we have $N$ particles distributed in the space. Since the free energy
is an extensive quantity, the total free energy of a DPD system can be
obviously expressed in terms of the free energy per DPD particle, $\psi$, as

\begin{equation}
{\cal F}=\sum_{i=1}^N \psi(n_i)=\sum_{i=1}^N \int d{\bf r} \delta({\bf r}-%
{\bf r}_i) \psi(n({\bf r}))=\int d{\bf r} \rho({\bf r} )\psi(n({\bf r})))
\label{free_part}
\end{equation}
where we have introduced the symbol $n_i$ to refer to the generalized
density defined above, although now expressed in terms of the positions of
the discrete $N$ DPD particles (see below). Comparing eqs.(\ref{free_part})
and (\ref{freevol}), we can easily identify $\psi(\rho)=f(\rho)$ which
obviously implies that we can decompose $\psi$ into its ideal and excess
contributions.

If the free energy determines the relevant energy for a given configuration
of DPD particles, we can then derive the force acting on each particle as
the variation of such an energy when the corresponding particle is
displaced. However, the motion of the particles themselves, due to the
action of the dissipative and random forces,  already accounts for the ideal
contribution to the free energy of the system, which is not related to the
interactions among the particles. Therefore, only the excess part of the
free energy will be involved in the effective interactions between the DPD
particles. Accordingly, we can write the conservative force acting on
particle $i$, ${\bf F}^C_i$, as

\begin{equation}
{\bf F}_i=-\frac{\partial}{\partial{\bf r}_i}\sum_{j=1}^N \psi^{ex}(n_j)
\label{force}
\end{equation}
We have derived the generic form for the conservative force acting on a DPD
particle as a function of the excess free energy that characterizes the
system, which is in general not pair-wise additive. These forces are
analogous to the ones derived from semi-empirical potentials\cite{FS} in MD,
used to model effectively the many-body interactions in condensed systems.
However, we have started from the macroscopic properties of the system, i.e.
its free energy, rather than ensuring microscopic consistency.

We can then fix the equilibrium thermodynamic properties of the system
beforehand, and derive a set of conservative forces consistent with the
desired equilibrium macroscopic behavior. This procedure is reminiscent of
an approach used in other mesoscopic simulation techniques that deal with
generic non-ideal fluids\cite{JY}.

Given that the free energy has been defined as a functional of a certain
local density, local variation in such a density are responsible for the
effective forces among the DPD particles. The particular expression for the
forces will then depend both on the specific form of the free energy and on
the choice of the local density $n_i$. It seems natural to define the local
density of a particle $i$ as its average on the corresponding interaction
range. For simplicity, we weight this average with the same functions used
to define the dissipative and random forces, as introduced in eq.(\ref
{eqts-DPD}). Therefore, we write 
\begin{equation}
n_{i}=\frac{1}{[w]}\sum_{j}w(r_{ij})  \label{dens}
\end{equation}
\noindent where $[A]$ refers to the spatial integral of a given quantity $A$%
. The normalization factor $[w]$ ensures that $n_{i}$ is indeed a density, so that in a homogeneous region, $n=\rho$.
This is in spirit similar to the weighted density approximation in density
functional theory\cite{Evans}. The use of a continuous and smooth weight
function that vanishes at the cut-off distance, $r_c$, ensures a smooth
sampling of the environment of each particle, avoiding spurious jumps. There
is no {\em a priori} reason to choose $w(r)$ equal to any of the other
weight functions, although the particular case of a
constant weight function constitutes a pathological limit - in this case the
conservative force will only act when one particle enters or leaves the
interaction range. The dependence of the energy of a particular
configuration on the particles' positions enters implicitly through the
weighted densities. For densities of the form given by eq.(\ref{dens}), the
conservative force acting on particle $i$ can be rewritten as

\begin{equation}
{\bf F}_{i}=-\sum_{j=1}^{N}\frac{\partial \psi(n_j)}{ \partial {\bf r}_{i}}%
=-\sum_{j}(\psi _{i}^{\prime }+\psi _{j}^{\prime }) \frac{w_{ij}^{\prime }}{%
[w]}{\bf e}_{ij}\equiv \sum_{j}{\bf F}_{ij}  \label{pairwise}
\end{equation}
where we have introduced the notation, $\psi_i\equiv\psi^{ex}(n_i)$, and
where the primes denote derivatives with respect to the corresponding
variables. Although the free energy of each particle depends on the local
density, and leads in general to many-body effective forces, for the
particular local density introduced in eq.(\ref{dens}), the forces between
DPD particles can still be written down as additive pairwise forces- a
computational advantage.

The fact that the forces depend on the positions of many particles through
their corresponding local weighted densities suggests that in general the
local structure of the fluid phase will be smoother than in the case in
which forces are derived from a pair-potential. This is an attractive
feature of the present model; the local structure in a fluid should only be
related to its microscopic structure, and should be smeared out at
mesoscopic, coarse-grained,  scales. In this respect, the density-dependent
interactions of these DPD models enforce an appropriate length scale
separation. In the next sections we will analyze these properties in detail.

Before considering specific examples, as a consistency check, we will
analyze the predictions for the pressure of a fluid following the free
energy, $p^{th}$, and the virial, $p^{v}$, routes. If we start from the free
energy per particle, eq.(\ref{free_part}), the pressure for a fluid will be

\begin{equation}
p^{th}=-f+\rho\frac{\partial f}{\partial \rho}=k_BT\rho+ \rho^2 \frac{
\partial \psi^{ex}}{\partial \rho}  \label{p:thermo}
\end{equation}

On the other hand, since we have derived the force between particles from
the free energy, we can also obtain the pressure of the fluid following the
virial route. In this case the pressure is given

\begin{equation}
p^{virial}=\rho k_{B}T+\frac{1}{2dV}\sum_{i}\sum_{j}{\bf r}_{ij}\cdot {\bf F}
_{ij}=\rho k_{B}T+\frac{1}{2dV}\int \int d{\bf r}d{\bf r}^{\prime }\rho ( 
{\bf r},{\bf r}^{\prime })({\bf r}-{\bf r}^{\prime })\cdot {\bf F}( 
{\bf r}-{\bf r}^{\prime })
\end{equation}
where we have approximated the discrete sum over the $N$ DPD particles by an
integral. Introducing the pair correlation function, $g(r)$, we can rewrite
the previous equation as 
\begin{equation}
p^{virial}=k_{B}T\rho +\frac{\rho ^{2}}{2d}\int d{\bf r}g(r)\frac{\partial
\psi ^{ex}}{\partial \rho }{\bf r}\cdot \left\{ \frac{-2w^{\prime }(r) {\bf e%
}}{[w]}\right\} =k_{B}T\rho -\frac{\rho ^{2}}{d}\frac{\partial \psi ^{ex}}{%
\partial \rho }\frac{[rw^{\prime }]}{[w]}  \label{p:virial}
\end{equation}
In the last equality we have assumed that the density is nearly homogeneous,
and that therefore $\partial \psi ^{ex}/\partial \rho $ is effectively a
constant. Otherwise, it is not possible to express the force in terms of the
relative coordinates only. If there is no local structure in the fluid, and $%
w(r_c)=0$, then $[rw^{\prime }]=-d[w]$, and then eq.(\ref{p:virial})
coincides with the prediction for the ``thermodynamic'' pressure, eq.(\ref
{p:thermo}) for any weight function\cite{note}. Otherwise,
a discrepancy between the two pressures will appear because the averaged
density $n_i$ is always centered on the corresponding DPD particle- a
conditional density- and it is therefore related to the $g(r)$. We will see
in some examples in subsequent sections how such local structure may develop.

Theoretical studies have shown that in the fluid phase of DPD, in the
hydrodynamic limit the usual Navier-Stokes equation is recovered\cite{Colin}%
, and that the equilibrium pressure term is related to the pairwise forces
through the usual virial expression, as we have derived previously. This
corresponds to dynamics which conserves momentum locally  (as in model-H\cite
{H-H}), instead of being purely relaxational (as happens in certain
dynamical models that start from density functional theories\cite{Marini}).
By analogy with the usual non-ideal DPD models, in equilibrium we recover a
probability distribution for a given configuration in agreement with
Boltzmann fluctuation theorem: the probability of observing a fluctuation is
proportional to the exponential of the deviation of the appropriate
thermodynamic potential- the free energy (as introduced in eq.\ref{freevol})
for DPD models at constant volume, temperature and number of particles.

In the following subsections we will consider three particular examples,
where we will compute explicitly the form of the conservative forces.

\subsection{Groot and Warren fluid}

\label{subsec:gw} Let us first derive the expression for the conservative
force that corresponds to the non-ideal fluid studied by Groot and Warren
\cite{Groot}. They introduce a conservative force of the form

\begin{equation}
{\bf F}_{ij} = \left\{ 
\begin{array}{cc}
a \left(1-\frac{r_{ij}}{r_c}\right){\bf e}_{ij} & ,r_{ij}<r_c \\ 
0 & ,r_{ij}>r_c
\end{array}
\right.  \label{GW_force}
\end{equation}

For this conservative force, they have shown that the EOS is $%
p=k_BT\rho+\alpha a\rho^2$, where by a numerical fit they found $%
\alpha=0.101\pm0.001$. Using the expressions of the previous section, the
corresponding pairwise force is 
\begin{equation}
{\bf F}_{ij} = \left\{ 
\begin{array}{cc}
2 \alpha a \frac{w_{ij}^{\prime}}{[w]}{\bf e}_{ij} & ,r_{ij}<r_c \\ 
0 & ,r_{ij}>r_c
\end{array}
\right.  \label{EOSGW}
\end{equation}
It corresponds to an excess free energy per particle $\psi^{ex} =\alpha a
\rho$, which is linear in the density. As stated in the introduction, an
interaction with a smooth, monotonic dependence in position does not induce
a fluid-fluid phase separation.

\subsection{van der Waals fluid}

\begin{figure}[ht]
\includegraphics[width=8cm]{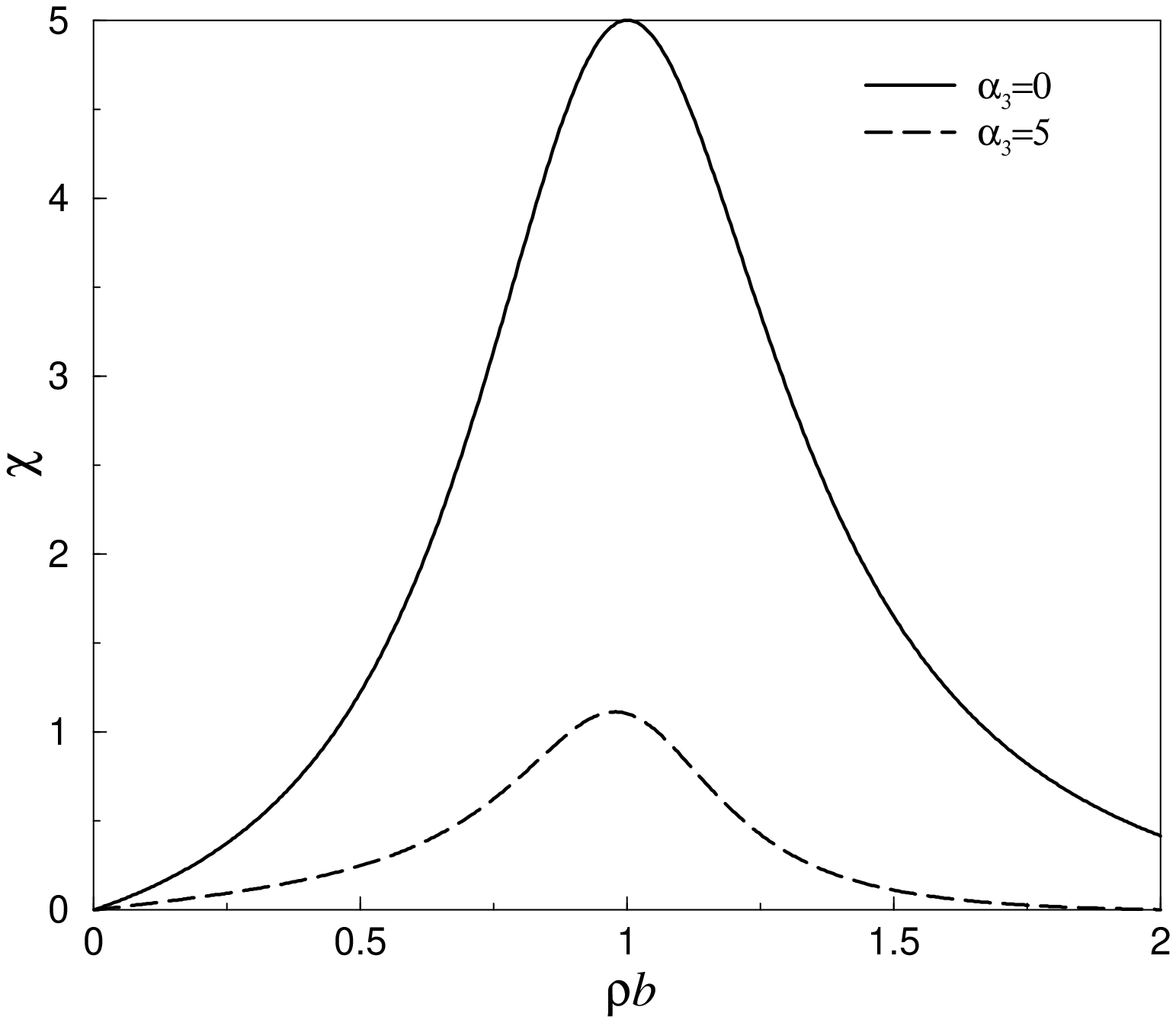}
\includegraphics[width=8cm]{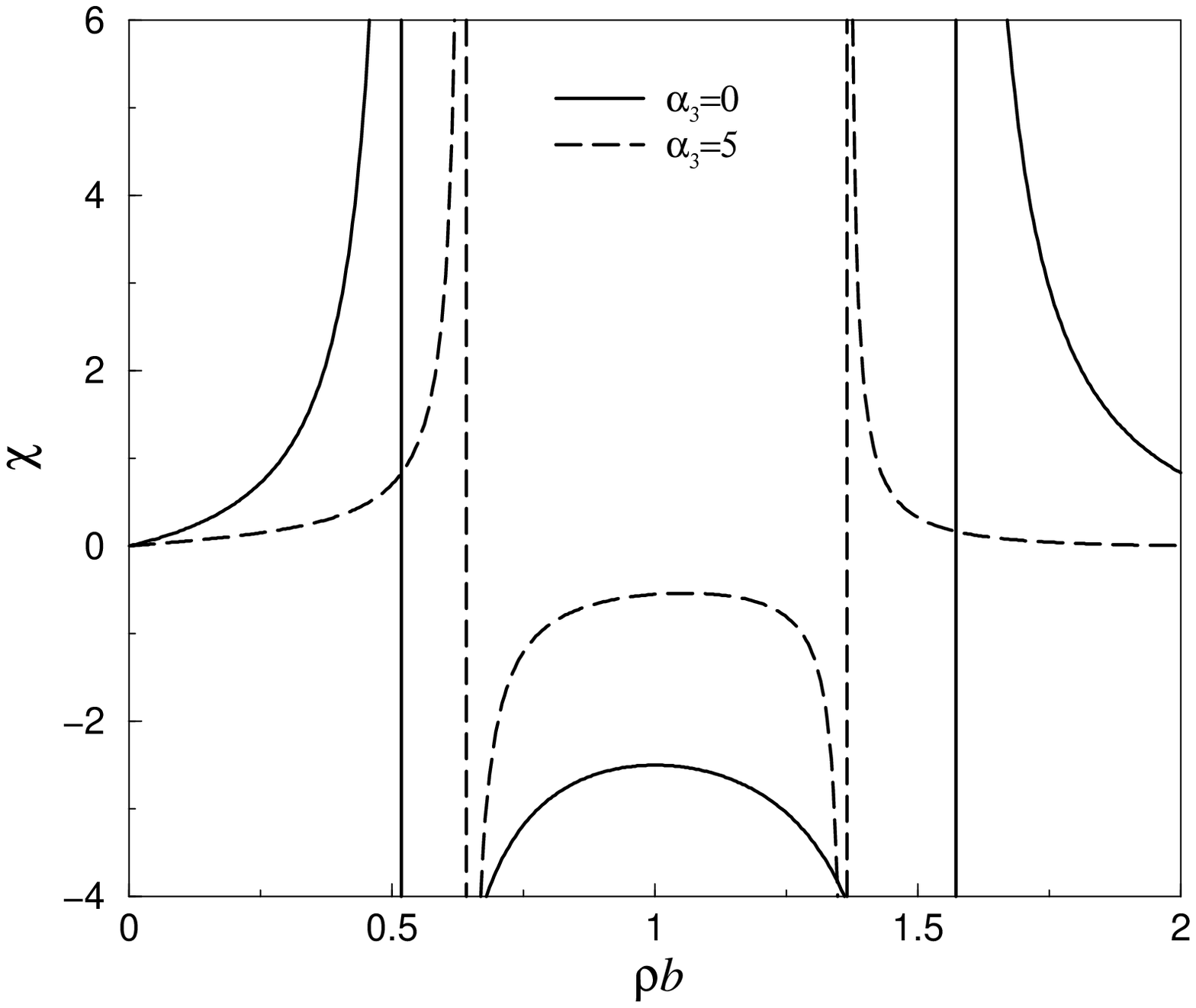}
\caption{Compressibilities of the van der Waals fluid around the critical
point, for two different values of the parameter $\protect\alpha_3 $. a)
Curves at $\protect T/T_{c}=1.1$; b) Curves at $\protect T/T_{c}=0.8$. }
\label{fig:xi}
\end{figure}

The van der Waals fluid is the classic example of a fluid with a liquid-gas
phase transition. It is characterized by the equation of state $p=\rho
k_{B}T/(1-b\rho )-a\rho ^{2}$ (and excess free energy per particle, $%
\psi^{ex} = -k_BT\log(1-b\rho)-a\rho$). We can recover this EOS in a DPD
system with pairwise conservative forces of the form,

\begin{equation}
{\bf F}_{ij} = \left\{\left(\frac{k_BT b}{1-b n_i}-a\right)+\left(\frac{k_BT b}{1-b n_j}-a\right)\right\}\frac{w^{\prime}_{ij}}{[w]} {\bf e}_{ij}
\label{EOSvdW}
\end{equation}
For reasons that will be discussed below, it is helpful to generalize
slightly the van der Waals fluid allowing for a contribution cubic in the
density. The EOS then becomes $p=\rho k_BT/(1-b \rho)-a \rho^2-\alpha_3 a b
\rho^3$. The critical point of this model corresponds to the parameters

\begin{eqnarray}
T_c&=& \frac{a}{b} b\rho_c(2+3\alpha_3 b\rho_c) (1-b\rho_c)^2  \nonumber \\
\rho_c &=&\frac{1}{b}\frac{\alpha_3-1+\sqrt{1+\frac{2}{3}\alpha_3+\alpha_3^2}
}{4\alpha_3}
\end{eqnarray}
\begin{eqnarray}
\rho_c b&\equiv&x_c = \frac{-1+\alpha_3+\sqrt{1+\frac{2}{3}%
\alpha_3+\alpha_3^2}}{ 4\alpha_3} \\
T_c b/a&\equiv& y_c = x_c(2+3\alpha_3 x_c) (1-x_c)^2
\end{eqnarray}
The compressibility of the fluid, $\chi$, in turn, can be written down as 
\begin{equation}
\chi^{-1} = \frac{k_BT}{\rho}+\frac{k_BT b (2-b \rho)}{(1-b\rho)^2}
-2a-3\alpha_3 a b \rho = \frac{y y_c}{x x_c (1-x x_c)^2}-2-3 \alpha_3 x x_c
\end{equation}
In fig.\ref{fig:xi} we show the behavior of the compressibility for two
different values of the parameter $\alpha_3 $, for temperatures close to the
critical temperature $T_{c}$. The increase in $\alpha_3$ reduces $\chi$ both
above and below the critical temperature. As expected, $\chi$ becomes
negative in a region below $T_{c}$ that is bounded by a spinodal.

Controlling the compressibility of the fluid is a desirable feature; a low
compressibility helps reducing fluctuations of the fluid interface, which
may be  useful in simulations. It also
provides a way of modifying properties of the fluid, such as the speed of
sound. Moreover, it gives an additional parameter to select the surface
tension which, as we will explain, may even change sign in this DPD-van der
Waals fluid. Finally, it proves useful to reduce the amplitude of the
density fluctuations to compare with mean field theoretical predictions, as
the ones developed in the next section.

\subsection{Binary mixture}

A binary mixture composed of particles of two species\cite{phasesep}, $A$ and $B$, has also
been considered by Groot and Warren\cite{Groot}. In this system,
it is possible to induce demixing with usual pairwise forces by modifying
the relative repulsions between the $A-A$, $B-B$ and $B-A$ pairs.
Nevertheless, even in this case, a model in which the forces depend on local
densities can be useful since if they induce less local structure, a
relevant feature at a fluid-fluid interface.

If the system consists of $N_{A}$ particles of type $A$ and $N_{B}$
particles of type $B$, then there are two relevant local density fields, $%
n_{A}$ and $n_{B}$, that are the straightforward generalizations of eq.(\ref
{dens}),

\begin{eqnarray}
n_{A_i}&=&\sum_{j\in A}\frac{w(r_{ij})}{[w]} \\
n_{B_i}&=&\sum_{j\in B}\frac{w(r_{ij})}{[w]}
\end{eqnarray}
$n_{A_i}$ and $n_{B_i}$ represent the concentration of $A$ and $B$ particles
around particle $i$, respectively. Whenever it is appropriate, we will 
denote by $\rho_A$ and $\rho_B$ the continuum limit of the discrete 
densities $n_A$ and $n_B$, respectively.

The simplest free energy that leads to a miscibility gap has an excess free
energy of the form

\begin{equation}
{\cal F}^{ex}= \int d{\bf r} \left\{2 \lambda \rho_a({\bf r}) \rho_b(%
{\bf r})+ \lambda_A \rho_a({\bf r})^2 + \lambda_B \rho_B({\bf r})^2\right\}
= \left[\sum_{i\in A}\left(\lambda n_{B_i}+\lambda_A n_{A_i}\right)+
\sum_{i\in B}\left(\lambda n_{A_i}+\lambda_B n_{B_i}\right) \right]
\label{freeen}
\end{equation}
where the two sums run over particles of type $A$ and $B$, respectively. The
corresponding conservative force acting on particle $i$ can be written down
as 
\begin{equation}
{\bf F}_j= \left\{\sum_{i\in A}\left\{\lambda\sum_{k\in
B}+\lambda_A\sum_{k\in A}\right\}+ \sum_{i\in B}\left\{\lambda\sum_{k\in
A}+\lambda_B\sum_{k\in B}\right\}\right\}\left[w^{\prime}_{ik}{\bf e}_{ik}
(\delta_{ij}-\delta_{kj})\right]
\end{equation}
Although in this case with two averaged local densities the conservative
forces do not have the form of eq.(\ref{pairwise}), they can still be
expressed as pairwise additive forces, 
\begin{equation}
{\bf F}_{ij}=\left\{ 
\begin{array}{ll}
-2 \lambda_{A,B} w^{\prime}_{ij}{\bf e}_{ij} &\;\;\; , i\;j \mbox{ same type} \\ 
-2 \lambda w^{\prime}_{ij}{\bf e}_{ij} & \; \; \;, i\;j \mbox{ different
type}
\end{array}
\right.  \label{EOSmix}
\end{equation}
This fluid will be miscible at high temperatures, and below a critical
temperature $T_c$ a miscibility gap will develop. In terms of the parameters
of the free energy, eq.(\ref{freeen}), for a symmetric mixture $T_c$ is 
\begin{equation}
k_BT_c=\rho (\lambda-\lambda_A)\;\;\;\;,\;\;\;\;\;\;\frac{\rho_A}{%
\rho_A+\rho_B}|_c\equiv c_c=\frac{1}{2}  \label{eq:critbin}
\end{equation}

\section{Interfacial behavior}

\label{sect:inter} In this section we develop a mean field theory for the
interfacial properties for a non-ideal DPD fluid that gives some insight in
the meaning of the conservative forces for these DPD models. For
definiteness, we concentrate on the derivation of the surface tension, $%
\tilde{\gamma}$.

Since we are interested in the interfacial properties, we focus on the
excess free energy, and will not write down the ideal gas contribution,
which is local in the density and does not contribute to the interfacial
properties. We start from the continuum limit of the appropriate free
energy, and make an expansion in gradients. Therefore, we disregard
correlations in the positions between the particles, hence the mean field
character of the predictions of the present section.

\subsection{van der Waals fluid}

\label{sect:FvdW} For a van der Waals fluid we can express the continuum
free energy of the fluid, that corresponds to the conservative forces
introduced in eq.(\ref{EOSvdW}), as 
\begin{equation}
{\cal F}^{ex} =\int d{\bf r} \rho({\bf r}) \left(-k_BT\log(1-bn({\bf r}%
))-an( {\bf r})-\frac{\alpha_3}{2}a b n({\bf r})^2\right)  \label{free_dft}
\end{equation}
where $n({\bf r})$ is the continuum limit of eq.(\ref{dens}), namely, 
\begin{equation}
n({\bf r}) = \frac{1}{[w]}\int d{\bf r}^{\prime}w(|{\bf r}-{\bf r}
^{\prime}|) \rho({\bf r}^{\prime})  \label{dens_cont}
\end{equation}
In eq.(\ref{free_dft}), the density $\rho({\bf r})$ means the mean density
at point ${\bf r}$. This is different from the density appearing in section 
\ref{sect:model}, where it referred to the instantaneous value of the
density for a particular configuration. Due to this density preaveraging,
the results of the present section constitute a mean field approximation.

For a smooth planar interface, we can expand the density in eq.(\ref
{dens_cont}) to second order in the gradients\cite{Evans}, 
\begin{equation}
\rho({\bf r}-{\bf z})=\rho({\bf r})-{\bf z}\cdot{\bf \nabla}\rho({\bf r})+ 
\frac{1}{2}{\bf z}{\bf z}:{\bf \nabla}{\bf \nabla}\rho({\bf r})
\label{dens_exp0}
\end{equation}
Inserting this expression in eq.(\ref{dens_cont}), and using the fact that
the weight function is radially symmetric we get 
\begin{equation}
n({\bf r}) = \rho({\bf r})+\frac{[z^2w]}{2 d[w]} \nabla^2\rho({\bf r})
\label{dens_exp1}
\end{equation}
With this expression, eq.(\ref{free_dft}) can be written down as 
\begin{eqnarray}
{\cal F}^{ex} &=& \int d{\bf r} \rho({\bf r})\left\{-k_BT\ln\left(1-b\rho(%
{\bf r})- \frac{b [z^2w]}{2 d[w]}\nabla^2\rho({\bf r})\right)-a\rho({\bf r})
\right.  \nonumber \\
&-&\left. \frac{a [z^2 w]}{2 d[w]}\nabla^2\rho({\bf r})-\frac{\alpha_3 a b }{%
2} \left(\rho({\bf r})^2+ \frac{[z^2 w]}{d[w]}\rho({\bf r})\nabla^2\rho({\bf %
r} )\right)\right\}  \label{free_exp1}
\end{eqnarray}
where terms containing derivatives higher than second order have been
neglected. Collecting terms in powers of the density gradients, making use
of the integration by parts we can rewrite eq.(\ref{free_exp1}) in the usual
form 
\begin{equation}
{\cal F}^{ex} = \int d{\bf r} \rho({\bf r}) \left(-k_BT\ln(1-b\rho({\bf r}
))- a\rho({\bf r})-\frac{\alpha_3}{2}a b \rho({\bf r})^2\right)+ \frac{[z^2
w]}{2 d[w]}\left(-\frac{k_BT b}{(1-b\rho({\bf r}))^2}+a+ 2\alpha_3 a b \rho( 
{\bf r})\right)|{\bf \nabla}\rho({\bf r})|^2  \label{free_exp2}
\end{equation}
The first term in brackets gives the local contribution to the excess free
energy. When the ideal contribution is added, it gives us the free energy
for a homogeneous van der Waals fluid. The second term in brackets is the
energy penalty to generate gradients in the system. It is this term that
contains, to lowest order, the interfacial energy of the fluid. In
particular, we can obtain from it an expression for the surface tension. If
we assume that the profile is a hyperbolic tangent, and we estimate its
width from the asymptotic bulk coexisting densities\cite{Godreche}, we
arrive at 
\begin{equation}
\tilde{\gamma}=\frac{\rho_l-\rho_g}{2}\sqrt{\frac{[z^2 w]}{d[w]}\left(-\frac{
k_BT b}{(1-b\rho_m)^2} +a+ 2\alpha_3 a b \rho_m\right) \frac{d^2 f}{d \rho^2}
}
\end{equation}
where $\frac{d^2 f}{d \rho^2}=1/\rho-2 a+k_BT (2-b \rho)/(1-b \rho)^2
-3\alpha_3 a b \rho$ is the second derivative of the homogeneous free energy
with respect to the density evaluated at one of the coexisting phases. We
have assumed for simplicity that the density difference between the two
phases is small, so that we can approximate the density across the interface
by its mean value, $\rho_{m}$

If we look at the structure of both the expansion of the free energy and the
surface tension, we can recognize a qualitative difference with respect to
the corresponding expressions for the standard van der Waals fluid. In the
latter, the interfacial tension is a function only of the parameter $a$
characterizing the long range attraction between the particles, whereas now
it depends on all the parameters, $a$, $b$ and $\alpha_3$. This qualitative
difference can already be traced back to the coefficient of the gradient
square term in free energy expansion, eq.(\ref{free_exp2}) - for the
standard van der Waals fluid the gradient energy cost is only related to $a$. As a result, in this DPD van der
Waals fluid there are different contributions to the gradient energy term
with different signs. Therefore, depending on their relative strength, it is
possible either to favor or penalize the appearance of density gradients in
the fluid; hence, the sign of the interfacial tension may change.

In an atomic fluid, the repulsion parameter, $b$, in the van der Waals EOS
arises from the hard core repulsion, while the attraction parameter, $a$,
comes from a long range weak attraction. Therefore, they appear in different
length scales, and accordingly, only the parameter $a$- related to the
long-range structure- is responsible for the behavior the interfacial tension. 
On the contrary, for a DPD
fluid there is no excluded volume interaction, and all interactions between
the particles take place at the same length scale, $r_c$. Then, the relative
strength of the different contributions will determine their overall net
effect. It is known that a microscopic model in which both attractions and
repulsions are long ranged leads to a van der Waals equation of state in
which the interfacial behavior can either favor or penalize the presence of
interfaces\cite{Sear}. The van der Waals fluid introduced in this paper
shares these same properties. Even if we can ensure a van der Waals EOS for
a fluid, a careful tuning of the parameters $a$, $b$ and $\alpha_3$ may lead
to a van der Waals model for lamellar fluid, whenever interfaces are
favored. Although unrealistic for atomic fluids, this behavior is relevant,
e.g for nanoparticles, for which repulsive and attractive interactions act
on similar length scales\cite{Sear1}.

\begin{figure}[ht]
\includegraphics[width=12cm]{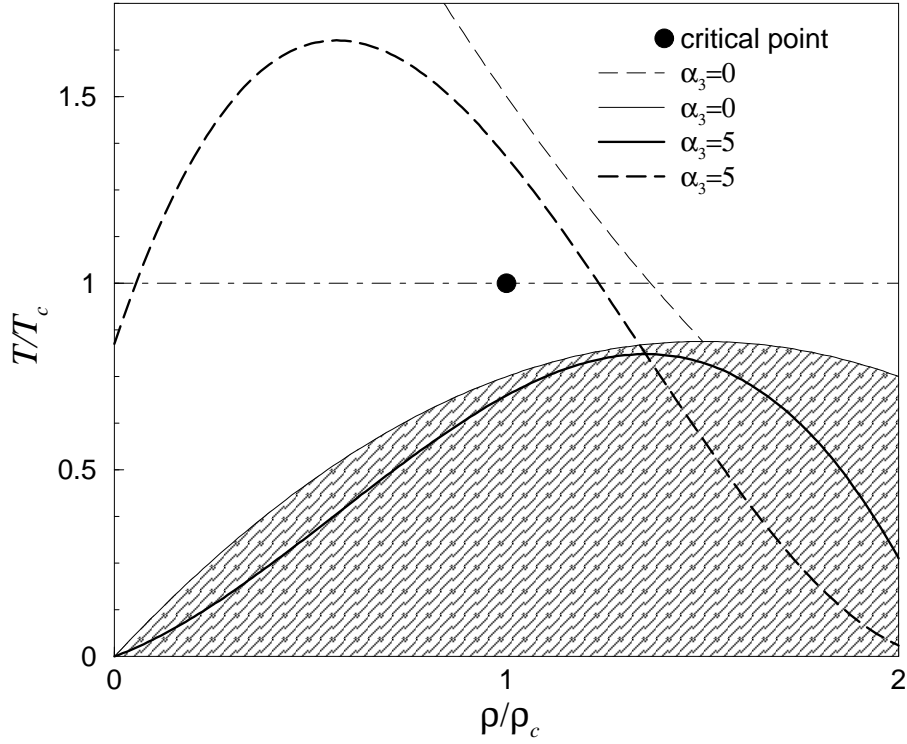}
\caption{Curves where the pressure and the surface tension vanish for two 
different values of $\protect\alpha_3$ for a van der Waals fluid. Above the 
solid curve the pressure is positive, and below the long dashed curves the 
surface tension is positive. The region contained in between the corresponding 
pair of curves corresponds to the portion of phase space where the fluid will 
be mechanically stable, with a positive surface tension. Above the long dashed 
curves the surface tension is negative. Two different values 
of $\protect\alpha_3$ are considered: $\protect\alpha_3=0$ 
and $\protect\alpha_3=5$.}
\label{fig:stability}
\end{figure}

Therefore, depending on the kind of fluid that needs to be modeled at
mesoscopic scales, the parameters in the free energy should be chosen
appropriately. For example, in order to get a positive surface tension, 
the densities of the fluid phases is restricted because one must ensure that both 
the pressure and the surface tension are positive.
In fig.\ref{fig:stability} we display the curves where the pressure and the 
surface tension vanish for two different values of $\alpha_3$. 
The area defined in between the corresponding set of curves defines the region 
of phase space where the fluid is mechanically stable with a positive surface 
tension. Remember that the values of $a$ and $b$ set the critical values  $\rho_c$ and $T_c$. The allowed regions do not change very much as the parameter $\alpha_3$ is modified.

If we make $b=0$, this model reduces to that of Groot and Warren. In
this case, $\tilde{\gamma}$ becomes negative (remember that $a$ is negative
now). As we have mentioned in subsection \ref{subsec:gw}, there is no
fluid-fluid phase separation in this model; therefore this negative value of
the surface tension does not lead to a proliferation of interfaces. However, 
the negative value of  $\tilde{\gamma}$ implies that the structure factor 
will have a minimum at a finite wave vector. We can define a characteristic 
length, $\tilde{l}_0$, on which local structure in the fluid will develop. 
If we  expand the free energy eq.(\ref{free_dft}) to  next order in gradients 
we can estimate this length to be
\begin{equation}
\tilde{l}_0 \sim 2\pi r_c \sqrt{\frac{[w r^4]}{12 [w r^2]}}
\end{equation}
which does not depend on the amplitude $a$; only on the shape of the weight
function $w$. Except for rapidly decaying weight functions, this length is 
of order of the interaction range $r_c$. This fact is consistent with the local 
structure observed in the radial distribution functions for this model
 (see section \ref{sec:gw}). We have also verified numerically the presence
 of a minimum in the structure factor $S(k)$.

\subsection{Binary mixture}
\label{sec:interbin}
We can also compute the interfacial tension for a binary mixture following
the procedure of the previous subsection. The excess free energy in the
continuum limit is now

\begin{equation}
{\cal F}^{ex}=\lambda \int d{\bf r} (\rho_A({\bf r}) n_B({\bf r})+ \rho_B( 
{\bf r}) n_A({\bf r}))+ \lambda_A \int d{\bf r} \rho_A({\bf r}) n_A({\bf r})
+ \lambda_B \int d{\bf r} \rho_B({\bf r}) n_B({\bf r})
\end{equation}

It is useful to introduce the total density $\rho$ and the mole fraction $c$
of component A as the relevant variables. They are defined as usual,

\begin{eqnarray}
\rho_A &=& \rho c \\
\rho_B &=& \rho (1-c)
\end{eqnarray}

If we expand the local densities $n({\bf r})$ in the same way as in eq.(\ref
{dens_exp1}), we arrive at the square-gradient approximation for the free
energy,

\begin{eqnarray}
{\cal F}^{ex}&=&\int d{\bf r} 2 \lambda \rho_A({\bf r}) \rho_B({\bf r})
+\lambda_A\rho_A({\bf r})^2+ \lambda_B\rho_B({\bf r})^2+\frac{[z^2 w]}{2 d
[w]}\left\{\lambda \rho_A\nabla^2\rho_B+ \lambda \rho_B\nabla^2\rho_A
+\lambda_A \rho_A\nabla^2\rho_A+ \lambda_B\rho_B\nabla^2\rho_B\right\} 
\nonumber \\
&=& 2\lambda \rho^2 c (1-c)+\lambda_A\rho^2 c^2+\lambda_B \rho^2 (1-c)^2+ 
\frac{[z^2 w]}{2 d[w]}\rho^2 \left\{-\lambda c\nabla^2 c-\lambda
(1-c)\nabla^2 c+\lambda_A c\nabla^2c- \lambda_B(1-c)\nabla^2z\right\} 
\nonumber \\
&=& 2\lambda\rho^2 c(1-c)+\lambda_A\rho^2c^2+\lambda_B\rho^2(1-c)^2+\frac{
[z^2 w]}{2 d[w]}\rho^2 \left\{2\lambda-\lambda_A-\lambda_B)|\nabla
c|^2\right\}
\end{eqnarray}
Assuming that $\rho$ is constant, and for a symmetric mixture ($%
\lambda_A=\lambda_B$), we get

\begin{equation}
{\cal F}^{ex}=\rho^2\int d{\bf r} \left\{ 2 (\lambda-\lambda_A) c (1-c)+2 \frac{[z^2 w]}{2 d[w]} (\lambda-\lambda_A) |\nabla
c|^2\right\}
\end{equation}
Again, the interfacial tension can have either a positive or negative sign,
depending on the relative magnitudes of the $\lambda$ parameters. If $%
\lambda_A=\lambda_B=0$, and only the repulsion between the particles
belonging to different species is kept, then the surface tension has the
same sign as $\lambda$, as expected. 

The interfacial width $\xi$ can be obtained taking into account that 
the concentration profiles converges exponentially to its bulk value. 
This gives us $\xi^2 = 4\kappa/F''$, where $\kappa/2$ is the amplitude 
of the $|\nabla c|^2$ in the gradient expansion of the free energy, 
and $F''$ is the second order derivative of the free energy with 
respect to the concentration evaluated at its bulk coexisting value. In the symmetric case, we get

\begin{equation}
\xi^2 = \frac{ [z^2 \omega]}{[\omega]}\left(-1+\frac{T}{4 T_c c_{\infty} (1-c_{\infty})}\right)^{-1}
\label{eq:xibin}
\end{equation}
where $c_{\infty}$ is the value of the concentration in the bulk phase.
 The surface tension, $\gamma$, can be obtained integrating the diference 
between the  free energy profile and its bulk value. In the small gradient 
 limit, it reduces to

\begin{equation}
\gamma = \int_{-\infty}^{\infty} \rho\frac{[z^2\omega]}{2 d[\omega]}T_c
 |\nabla c|^2
\label{eq:gammaexp}
\end{equation}
If we assume that the concentration profile is a $\tanh$, we get the estimate

\begin{equation}
\gamma = \frac{2 [z^2 \omega]}{3 [\omega]}\frac{\rho T_c}{\xi}( c_{\infty}-1/2)^2
\label{eq:sigmabin}
\end{equation}
Close to the critical point, we recover the expected limiting behavior for the interfacial properties\cite{Godreche},

\begin{eqnarray}
\gamma &=&2 \rho
k_{B}T_{c}\frac{T_{c}}{T}\sqrt{\frac{2 [z^2\omega]}{3 d [\omega]}\left( 1-\frac{T}{T_{c}}\right)
^{3}} \\
\xi &=&\sqrt{\frac{[z^{2}w]}{4d[w](1-T/T_{c})}}
\end{eqnarray}

\section{Equilibrium properties}

\label{sect:comp} We will now analyze the equilibrium properties of the
examples of non-ideal DPD systems introduced in section \ref{sect:model} and
will compare with the predictions of previous models performing numerical
simulations. We  take the interaction range $r_c$ as the unit of length, the
mass of the DPD particles $m$ as the unit of mass, and the critical thermal
energy, $k_BT_c$, of the corresponding free energy as the unity of energy.
If no phase transition is present, then $k_BT$ is taken as the unit of
energy. The equations of motion are integrated self-consistently to avoid
spurious drifts in the thermodynamic properties\cite{IF}.

\subsection{Groot and Warren fluid}

\label{sec:gw} Before studying a DPD model with fluid-fluid coexistence, we
compare the results of our model for a Groot-Warren fluid with the original
one, based on forces given by eq.(\ref{GW_force}). In this case, both models
should coincide and we analyze it to see the effects of the weight function 
shape on the  properties of non-ideal fluids..

\begin{figure}[ht]
\includegraphics[width=12cm]{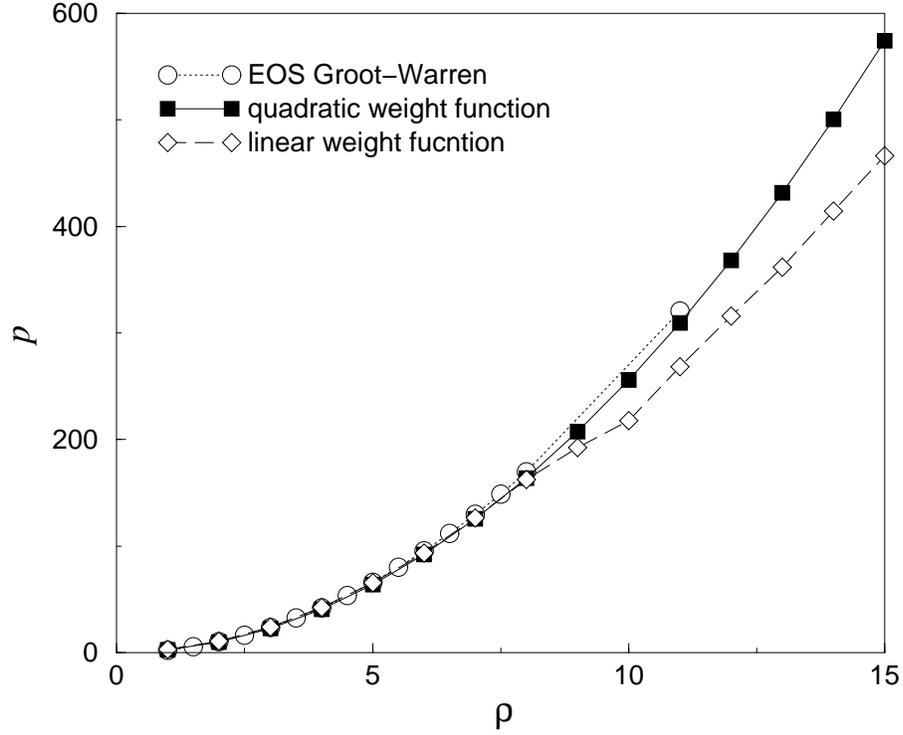}
\caption{Pressure as a function of the density for a Groot-Warren fluid,
using both the previously proposed pairwise force, eq.(\ref{GW_force}), and
for the force of the present form, eq.(\ref{EOSGW}). In the second case we
compare the behavior for a linear and a quadratic weight function. $a=25$, $%
\protect\alpha=0.101$. $L=6 r_c$, $k_BT=1 $, $\protect\gamma=1$ (See head of
sec.\ref{sect:comp} for units).}
\label{fig:GW}
\end{figure}

We have performed simulations for a DPD fluid in 3 dimensions, taking as
parameters $a=25$ and $\alpha =0.101$- which corresponds to those used in 
\cite{Groot}. In fig.\ref{fig:GW} we compare the predictions for the EOS
given by our model and by running a DPD simulation with the Groot-Warren
model.

Groot and Warren used the same weight function for all pairwise forces. The
proposed model for this non-ideal fluid shows neatly that for the present
class of models a linear weight function is not suitable to sample the local
density of each DPD particle, because it leads to a pairwise conservative
force that exhibits a discontinuity at the edge of the interaction region, $%
r_c$. We have analyzed the effect of such a jump on the thermodynamic and
structural properties of this system. To this end, we have considered both
decreasing linear and quadratic $w$'s.

\begin{figure}[ht]
\includegraphics[width=12cm]{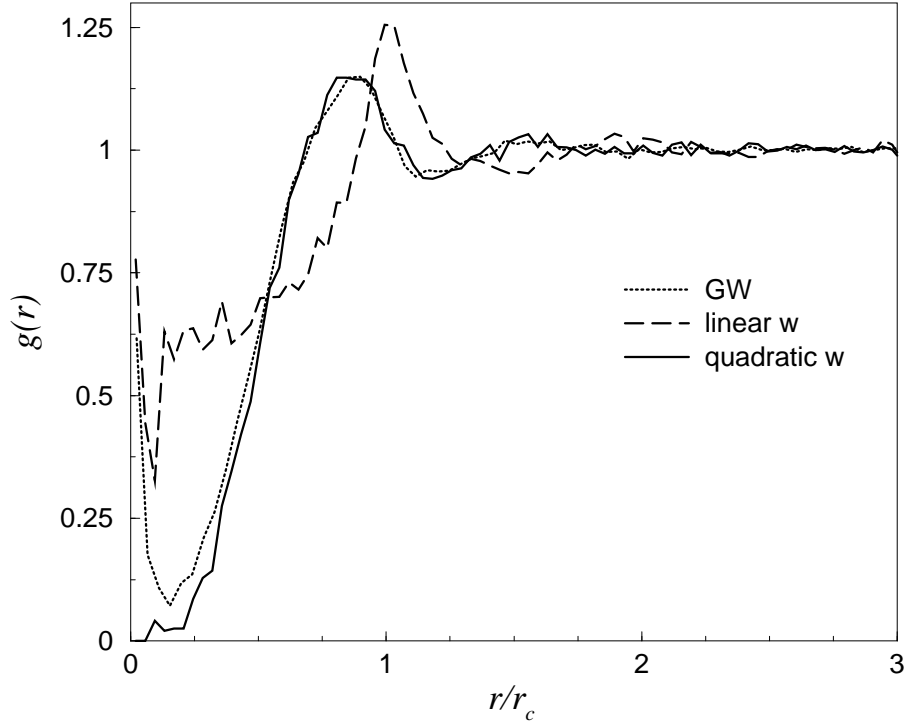}
\caption{Radial distribution for a Groot-Warren fluid, using both the
previously proposed pairwise force, eq.(\ref{GW_force}), and for the force
of the present form, eq.(\ref{EOSGW}). In the second case, we compare the
behavior for a linear and a quadratic weight function. Same parameters as in
fig.\ref{fig:GW}. The mean density is $\protect\rho_m=3$. }
\label{fig:gr_GW_n3}
\end{figure}

\begin{figure}[ht]
\includegraphics[width=8cm]{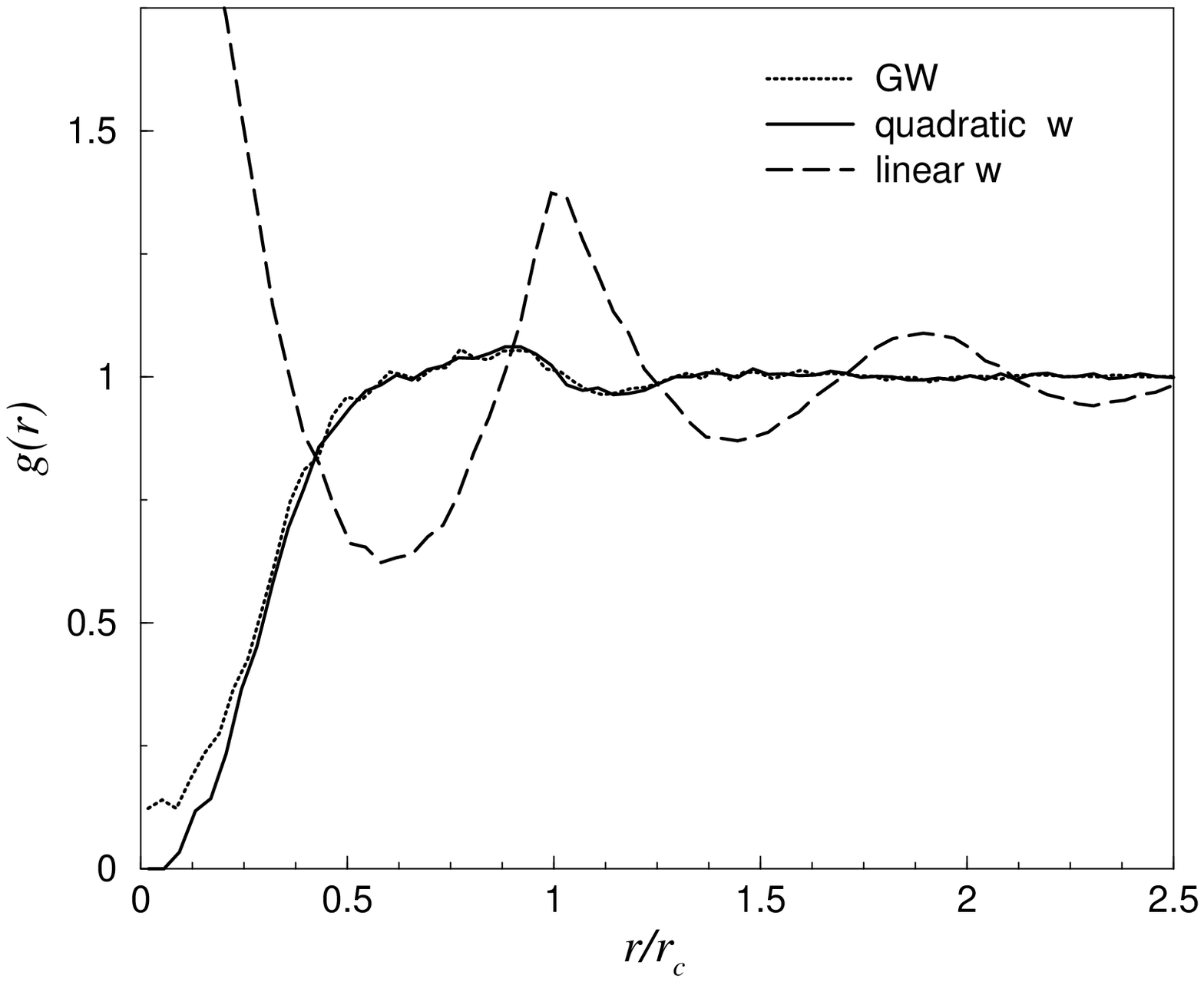}
\includegraphics[width=8cm]{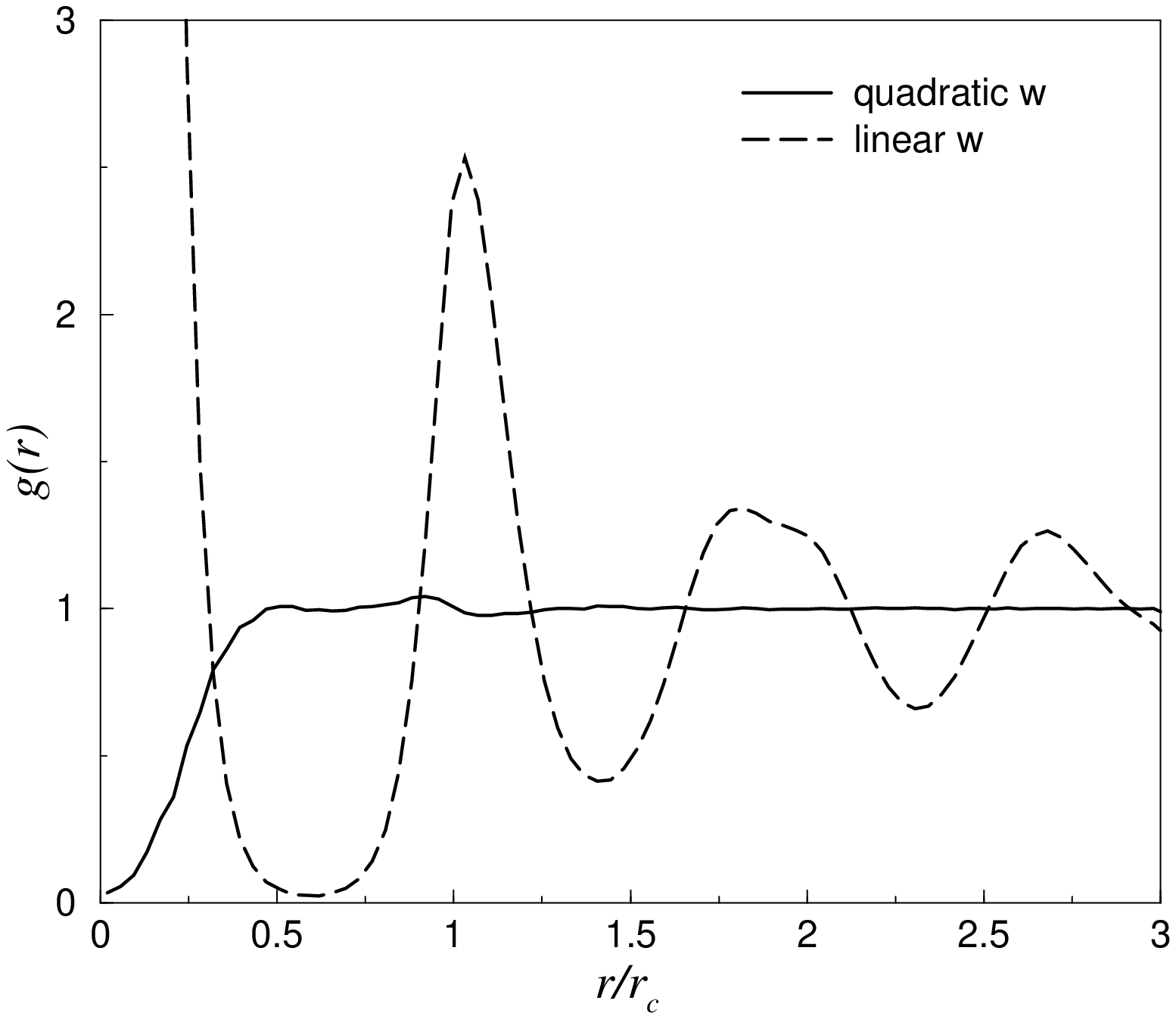}
\caption{Radial distribution function for a Groot-Warren fluid, using both
the previously proposed pairwise force, eq.(\ref{GW_force}), and for the
force of the present form, eq.(\ref{EOSGW}). In the second case we compare
the behavior for a linear and a quadratic weight function. Same parameters
than in fig.\ref{fig:GW}. The mean densities are: a) $\protect\rho_m=8$ and
b) $\protect\rho_m=14$.}
\label{fig:gr_GW_n8}
\end{figure}

In fig.\ref{fig:GW} we compare the EOS obtained from simulations; for a
quadratic $w$, our model coincides with that of Groot-Warren. However, for a
linear $w$, the agreement survives only at low densities. This DPD model has
a transition to a solid state at high densities, and the results obtained
indicate that the location of such a transition is sensitive to the shape of
the weight function - the characteristic force felt by each particle depends
on the shape of $w$ for a given density. In figs.\ref{fig:gr_GW_n3}-\ref
{fig:gr_GW_n8} we compare the radial distribution functions for our model
and that of Groot-Warren, and for for different $w$'s, at increasing values
of the density. It is  clear that the shape of $w$ plays an important role
in the local structure of the fluid, and will influence the location of the
fluid-solid transition. In section \ref{sect:FvdW} we have noted that for
the present model there exists a characteristic length, $\tilde{l}_0$,
associated with density fluctuations and which is of order $r_c$. Only for
fairly narrow weight functions will this length become much smaller than $r_c
$.

At low densities, a linear $w$ generates less local structure, a pleasant
feature for a mesoscopic model. However, as the density is increased, the
local structure develops faster for a linear weight function, leading 
sooner to a transition to the ordered phase. The use of a quadratic weight
function leads to results identical to those of the GW model, while a linear
force tends to smooth the structure at short distances. The mean repulsion
between particles is larger with a linear $w$ rather than with a quadratic
one. Moreover, it seems plausible to assume that the discontinuity in the
force induces a higher sensitivity to local density fluctuations. These
results show how the modifications of the shape of the weight function can
be used to tune fine details of the behavior of a fluid, once the EOS has
been fixed.

\subsection{van der Waals fluid}

\label{sec:lg}

\begin{figure}[ht]
\includegraphics[width=12cm]{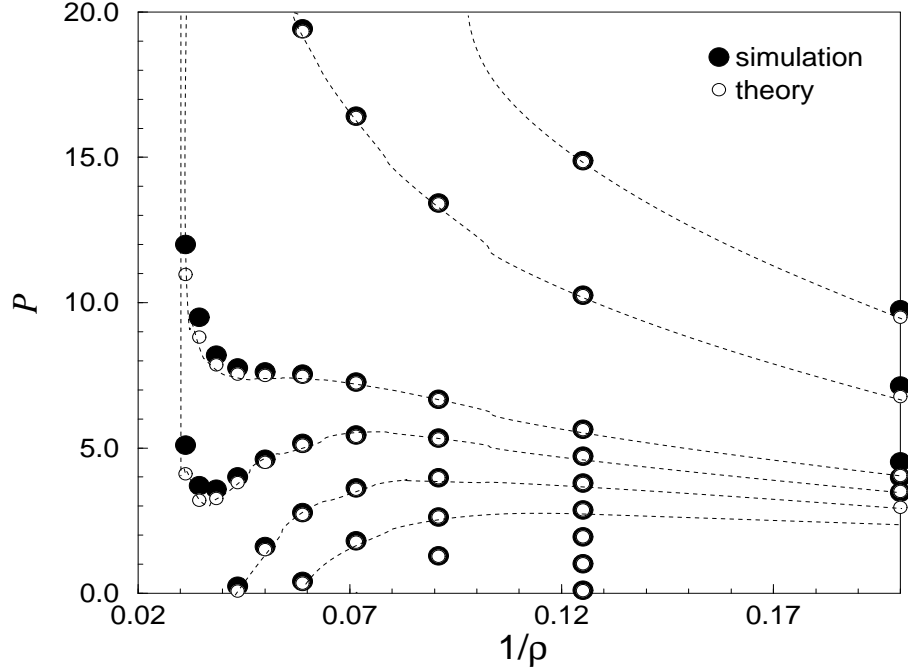}
\caption{ Equation of state for a 2-D van der Waals fluid. The different
sets of data points correspond to different temperatures. $b = 0.016$, $%
a=1.9 b$, $\protect\alpha_3 = 5 $, $L=7 r_c$, $\protect\gamma = 1$ (See head
of sec.\ref{sect:comp} for units).}
\label{fig:a3}
\end{figure}

Next, we focus on the liquid-gas equilibrium properties of a two-dimensional
van der Waals fluid. Taking a homogeneous system, we can analyze the
effect of the density fluctuations on the EOS, and compare it with the
predictions coming from the macroscopically assumed EOS. In fig.\ref{fig:a3}%
, we show the pressure values obtained in simulations run at fixed
homogeneous density, volume and temperature. In this case we can recover the
characteristic van der Waals loop. The actual coexistence curve should be
derived from it using the equal area Maxwell's construction. The agreement
with the expected EOS from the macroscopic free energy is very good, and
only small deviations are observed, due to particle correlations.

We have also analyzed the density and pressure profiles when we bring into
contact a liquid and a gas in the coexistence region. As mentioned in
section \ref{sect:model}, the compressibility of the fluid, especially in
the coexistence region, is very sensitive to the parameter $\alpha_3$ that
characterizes the amplitude of the term cubic in the pressure. For $%
\alpha_3=0$ the density profiles tend to fluctuate substantially. Note that
our estimates for the parameters and ranges of stability are all based on a
mean field description, which may be no longer quantitatively correct under
such conditions. Due to this, a  series of simulations will be needed for
each set of selected parameters whenever a detailed, quantitative
comparison, may be required.

When the parameter $\alpha_3$ is increased (we have taken the value $%
\alpha_3=5$), imposing an initial slab of liquid in coexistence with a slab
of gas the interface remains stable, and the density fluctuations in the
liquid phase are not too large.

\begin{figure}[ht]
\includegraphics[width=12cm]{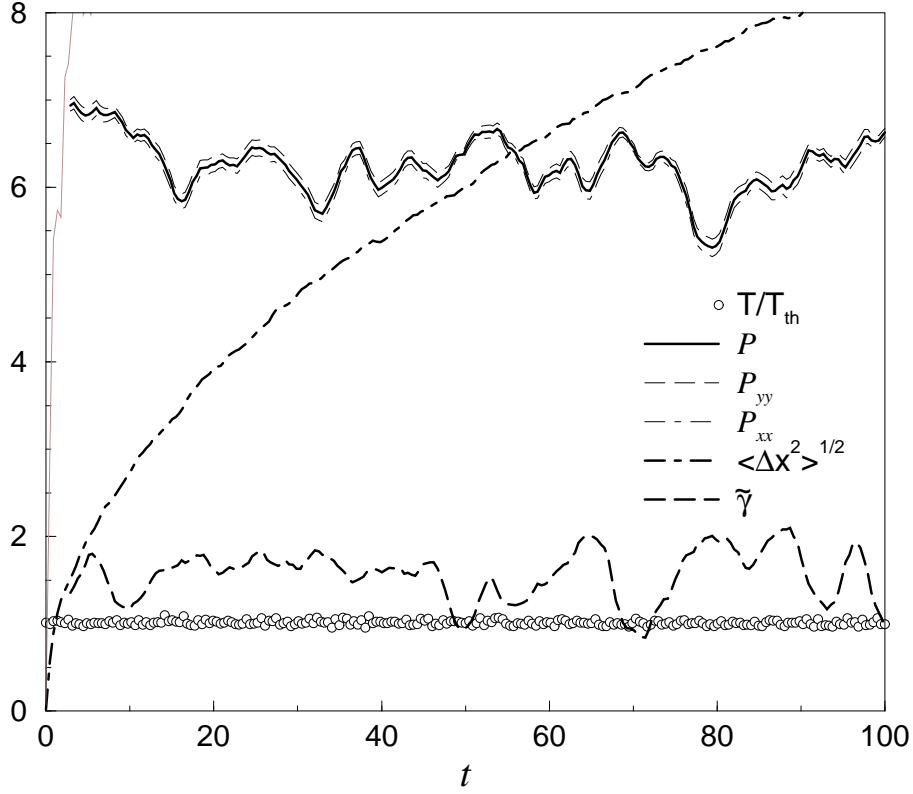}
\caption{ 
Thermodynamic values of a DPD fluid with a van der Waals EOS, when
a liquid is coexisting with the gas phase, in two dimensions. The initial
condition corresponds to a slab of fluid in the $\protect y$ direction in coexistence
with a slab of gas. $\tilde{\protect\gamma}$ is the interfacial tension,
extracted from the mean pressures, $\tilde{\protect\gamma}= (L_y/2)
(P_{yy}-P_{xx})$. Also displayed the mean square displacement in units of
the interaction range $r_c$. $L_y=20$, $L_x=3$, $k_BT=0.75$, $a=1.9*b$, $%
b=0.0156$, $\protect\alpha_3=5$. The unit of time is the time needed for a
DPD particle to diffuse $\protect r_c$ initially. (See head of sec.\ref{sect:comp}
for units). }
\label{fig:templg}
\end{figure}

In Fig.\ref{fig:templg} we show the temperature, pressure and mean square
displacement of the system during the extension of the simulation. One can
see that the temperature does not shift, and corresponds to its nominally
assigned value. The pressure exhibits important fluctuations, but if we
subtract the normal and tangential components (in the figure we only display
the averaged pressure), their difference, which is twice the surface
tension, gives a value with a well-defined positive mean. Also  the mean
square displacement shows that particles have had the time to diffuse the
interfacial width, which is roughly proportional to the interaction range, $%
r_{c}$, indicating that the droplet is stabilized.

\begin{figure}[ht]
\includegraphics[width=8cm]{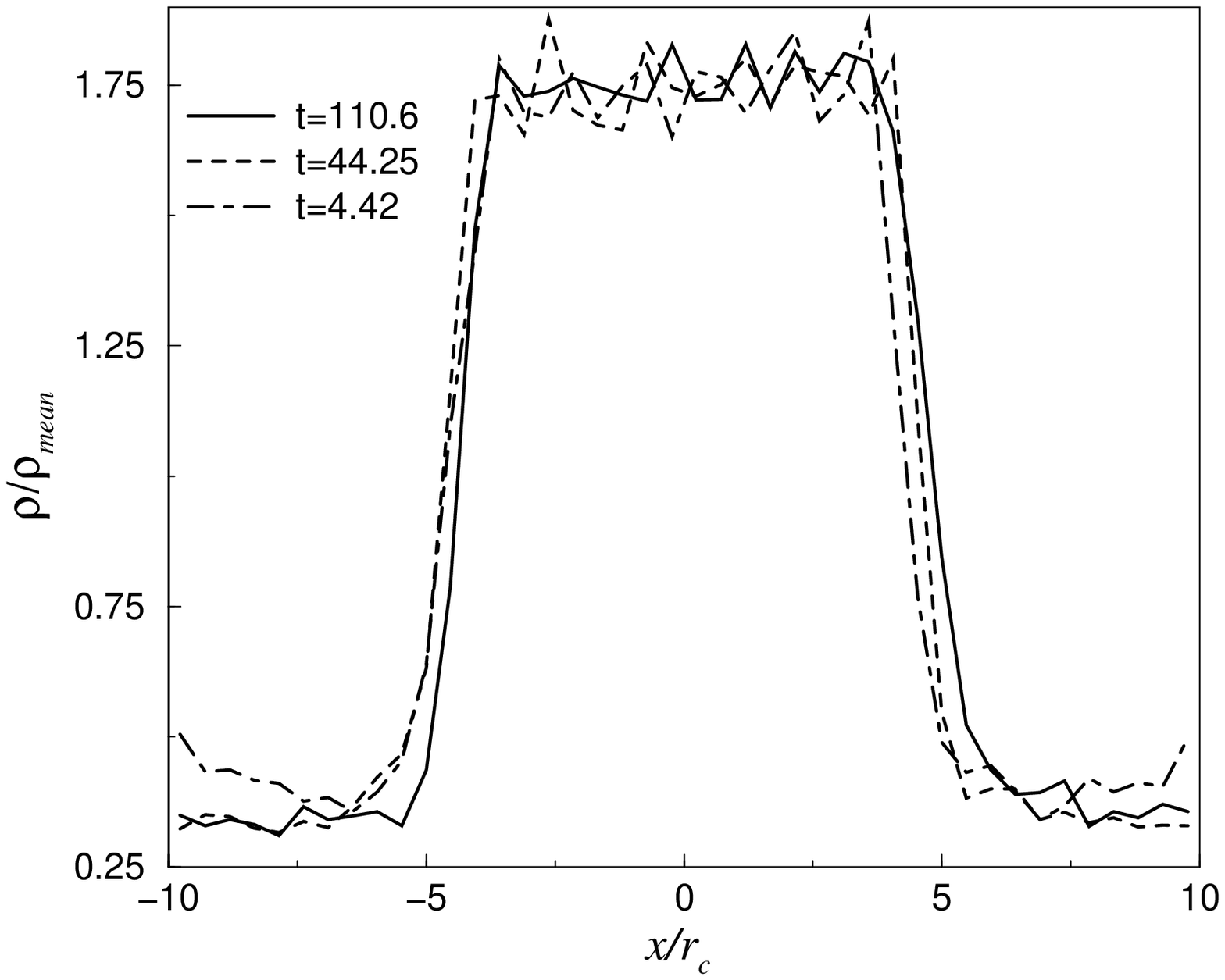}
\includegraphics[width=8cm]{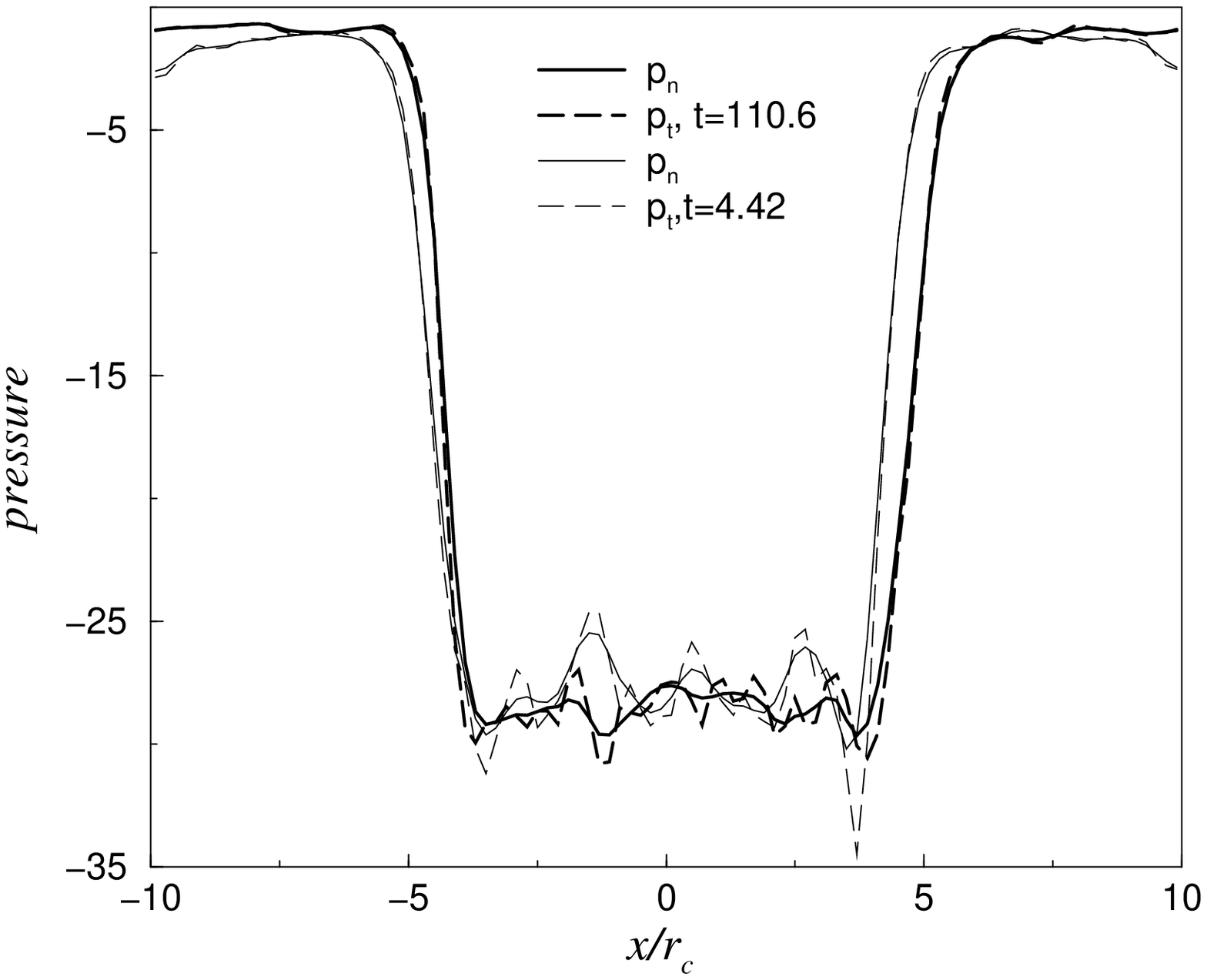}
\caption{ Equilibrium a) density and b) pressure profiles for a 2-D van der
Waals fluid. The initial profile is a step profile. Same parameters as in
fig.\ref{fig:templg} (See head of sec.\ref{sect:comp} for units).}
\label{fig:denslg}
\end{figure}

Fig.\ref{fig:denslg}a shows the density profiles obtained by starting with a
step density profile in the liquid-gas coexistence region, where the numerical errors are smaller than the fluctuations, as in the rest of the plots. The shape of the
drop is stable and the interfaces fluctuate around their initial location,
as could be expected. The density ratio between the two fluid phases, $%
\rho_{liq}/\rho_{gas}=4$ makes it reasonable to call the two phases  liquid
and gas. The density in the gas phase is $10 r_c^{-2}$, which ensures that
in both phases the number of interacting particles is sufficiently high. By
looking at the density profiles one can also observe that the density
fluctuations in the dense phase are small, as expected on the basis of the
small compressibility of the fluid.

\begin{figure}[ht]
\includegraphics[width=12cm]{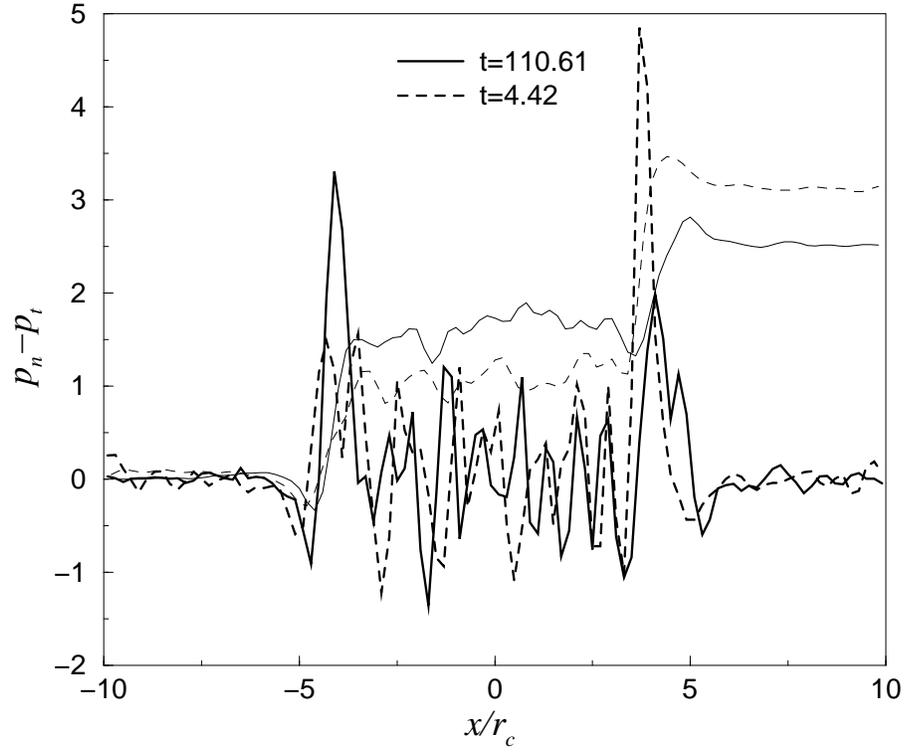}
\caption{ Profiles of the difference between the normal and tangential
components of the pressure tensor along the system, for the pressure
profiles of fig.\ref{fig:denslg}b. Same parameters as in fig.\ref{fig:templg}
(See head of sec.\ref{sect:comp} for units).}
\label{fig:diffpress}
\end{figure}

Finally, we have also computed the components of the pressure tensor across
the profiles. For an inhomogeneous fluid there is no unambiguous way of
computing the local components of the pressure tensor; we follow here the
procedure described in ref.\cite{Irving} and display them in fig.\ref
{fig:denslg}b. They follow basically the increase in density, exhibiting
larger fluctuations in the liquid phase. In the bulk phases, the two
components of the pressure tensor have to be equal. This is clearly shown in
fig.\ref{fig:diffpress}, where the differences in the two components are
confined to the interfaces, if we compare the location of the differences
with the density profiles of fig.\ref{fig:denslg}a. Moreover, the increase
in fluctuations in the dense phase is clearly displayed. The equilibration
of the drop can also be monitored by analyzing the time scale at which the
pressure profile becomes symmetric at both interfaces. Together with the
pressure differences, we have also plotted in thin line the integral of the
pressure difference across the profile. This quantity is the surface
tension, and indeed, the values we get when the profile is equilibrated
agree with the predictions extracted from the mean pressures, displayed in
fig.\ref{fig:templg}. We have also computed the excess free energy profile.
Its integral gives us an alternative (thermodynamic) route to compute the surface tension. We
have verified that the values of the surface tension obtained integrating
the excess free energy profile coincides with the value presented above,
computed along the virial route.

Another appealing feature of these conservative interactions is that their 
density dependence induces smooth local structure. Indeed, if we analyze the
radial distribution functions for a homogeneous phase, we can see that the
structure in this case is almost non-existent. When an interface is present
it is hard to assess the spurious structure that the model may induce
through the density profile . All we can say is that the decay of the
density is monotonic from one phase to the other, and avoids therefore
spurious structure close to the interface. Such a structure would be
spurious on the mesoscopic scale modeled by the DPD fluid. In contrast, the
onset of structuring of the liquid-vapor interface on an atomic scale
(beyond the Fisher-Widom line) is a real effect\cite{Evans2}.

\subsection{Binary mixture}

\label{sec:bin} 
\begin{figure}[ht]
\includegraphics[width=12cm]{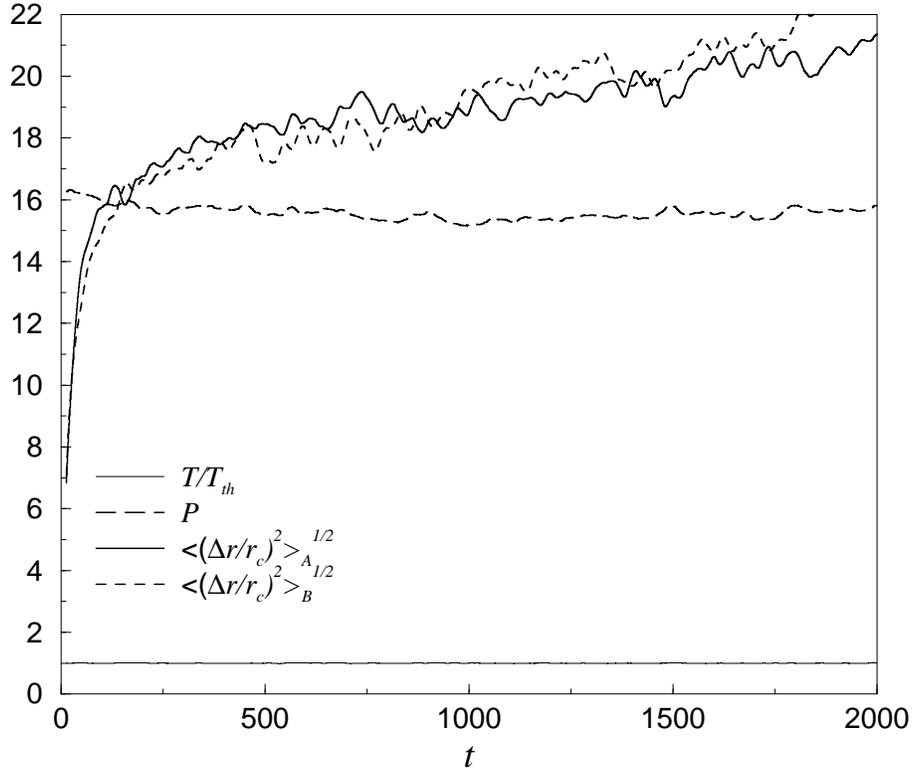}
\caption{ Temperature, pressure and mean square displacements of the two
species as a function of time, for a binary mixture below its critical
temperature, $T/T_c=0.5$, and with $\protect\lambda =1$, $\protect\lambda%
_A=0.2$ (See head of sec.\ref{sect:comp} for units). }
\label{fig:tempmix}
\end{figure}

Finally, we have run simulations for a binary mixture
corresponding to the model described in section \ref{sect:model}.C. As in
the previous subsection, we concentrate on the equilibrium properties of the
fluid in the coexistence region. We have simulated a 2D fluid, starting with
an initial step profile in concentration. In fig.\ref{fig:tempmix} we show
the evolution of the temperature and pressure, which remain essentially
constant through the simulation. We also display the root mean square
displacement of the two species. One can clearly see that after a short
initial period, when the species start to feel the presence of the
interfaces their effective diffusion slows down. The fact that the mean
square displacement is much larger than the interfacial width, which remains
of the order of the interaction range, $r_c$, ensures that the initial
configuration has relaxed to its proper equilibrium shape.

\begin{figure}[ht]
\includegraphics[width=12cm]{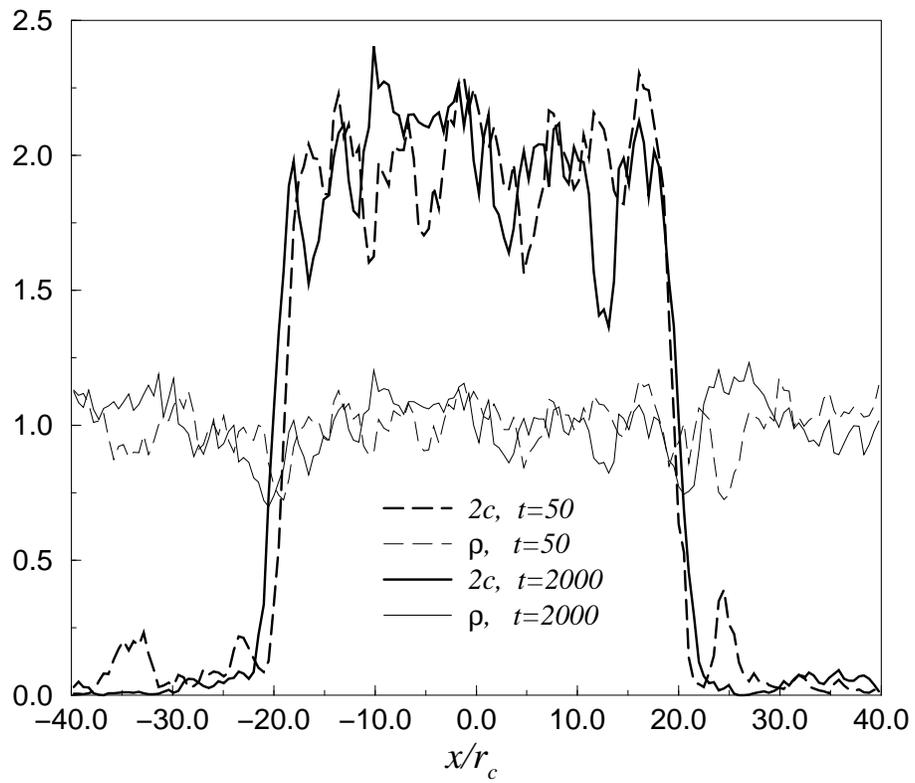}
\caption{ Profiles of the relative amount of one of the species across the
system, at two different times. These curves have been multiplied by $2$ to
avoid confusion with the thin lines. The latter correspond to the normalized
mean density at the same time (See head of sec.\ref{sect:comp} for units).}
\label{fig:densmix}
\end{figure}

We have computed the concentration profiles as a function of time. In fig.%
\ref{fig:densmix} we show the concentration profiles of one of the species
at an initial and late stage of the relaxation towards the equilibrium
coexistence. As was the case in the van der Waals fluid, the fluctuations
are greater in the concentrated phase. Although the concentration of each
species goes basically to zero in one of the two coexisting phases, the
interface does not broaden and keeps its width within $r_{c}$. Despite this
large concentration gradient, the mean density barely changes across the
interface. These normalized mean densities are displayed in also in 
fig.\ref{fig:densmix} as thin curves. Although a small dip in the normalized mean
density appears at the interfaces, its value is not large compared with the
typical bulk density fluctuations (which are due to the compressibility of
the fluid). Again, this indicates that the use of concentration dependent
conservative forces suppresses the appearance of spurious structure at
interfaces, while still being able to drive the phase separation.

\begin{figure}[ht]
\includegraphics[width=12cm]{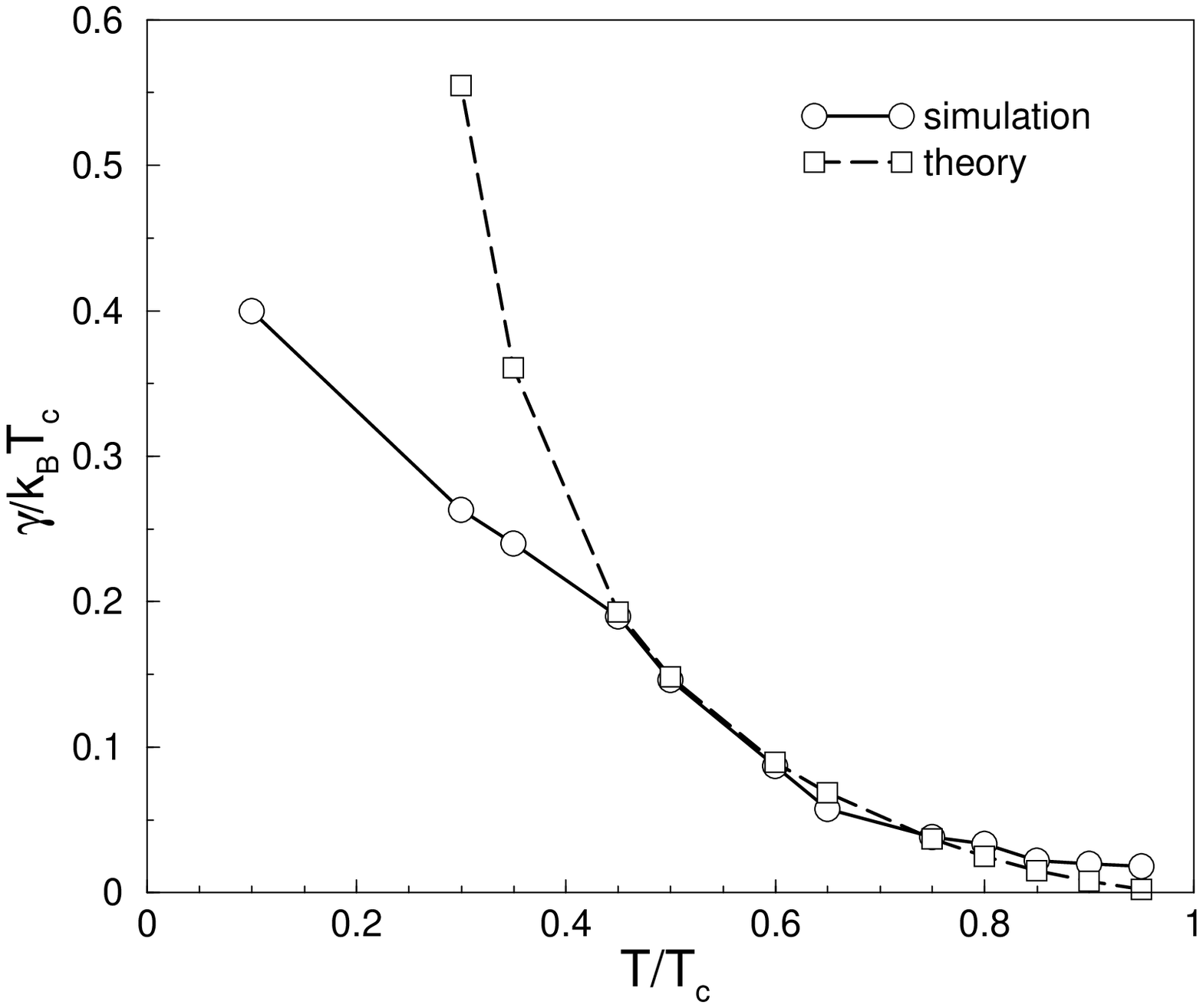}
\caption{Surface tension for a binary mixture  at density $\rho=0.5$ with a 
critical temperature $T_c=8$ and quicomposed, as a function of the temperature.
 The squares correspond to the expression derived from the mean field free 
energy in the small gradient limit. }
\label{fig:tension}
\end{figure}

We can also test the predictions of section
 \ref{sec:interbin} for the interfacial properties on the basis of 
 a binary mixture.  To this end, we have 
integrated numerically eq.(\ref{eq:gammaexp}) using the concentration 
profiles obtained from the simulations, and we have compared the results 
with the theoretical prediction, eq.(\ref{eq:sigmabin}). We display the results 
in fig.\ref{fig:tension}, where we have multiplied the theoretical curve by an 
overall numerical factor, since the numerical prefactors in eq.(\ref{eq:sigmabin}) 
are approximate. One can observe  that the overall good agreement is lost at small 
temperatures, where the interface is very sharp, and close to the critical point, 
where fluctuations are expected to play a relevant role.

\section{Conclusions}

\label{sec:disc}

We have presented a new way of implementing conservative forces between DPD
particles. Rather than assuming a force that depends on the interparticle
separation, we have introduced a conservative interaction that depends on
the local excess free energy. In this way, it is possible to fix beforehand,
at the mean field level, the desired thermodynamic properties of the system.
However, this procedure neglects the effect of particle correlations.
Whenever an accurate quantitative comparison is needed, a set of numerical
simulations will be required to determine accurately the appropriate phase 
diagram. We could equally use the free energy to carry out Monte Carlo 
simulations to analyze the static properties of fluids; this procedure 
will suffer from similar drawbacks as a result of the ignored particle 
correlations.

When the free energy per particle depends on the averaged local density, it
is possible to recover central pairwise additive forces- an important
computational feature.  The only assumption we have made is that the system
is isothermal, although it should be straightforward to generalize it to
include energy transport, along the lines developed previously\cite{Avalos}.

These models can be viewed as a dynamical density functional theory (DFT)
for smooth conservative forces with local momentum conservation. However,
since the DPD particles do not have a local structure, these models can only
describe the dynamics at a mesoscopic level, while the usual dynamical DFT
can account for the dynamics down to the microscopic scale. 

In addition to the freedom in the choice of the free energy, this new type
of proposed forces leads to weaker structure at short distances. Hence, we
can enforce a proper length and time scale separation, avoiding the
appearance of microscopic features of the system at distances of order $r_c$.

At the mean-field level, and using standard techniques, it is easy to derive
expressions for the interfacial properties. We have shown that the absence
of internal structure of the DPD particles (implying that all forces act on
the same length scale) leads to qualitatively new behavior not present in
atomic fluids. From the physical point of view it shows that, for example,
the same thermodynamic system can be tuned to favor macroscopic or
microscopic phase separation. Although it may seem unrealistic, the
competition of attractive and repulsive effective potentials on the same
length scales correspond to certain physical situations, and they are
probably more common on the mesoscopic than in the microscopic domain. In this
respect, the models we have introduced are quite flexible because, for a
given bulk thermodynamic behavior (e.g. a given EOS), it is still possible
to modify the parameters to control other physical properties. For example,
the mean interaction strength can be changed by modifying the way in which
the local density is sampled, or for the van der Waals fluid, it is possible
to modify the compressibility (and hence the speed of sound). As in any 
diffuse interface model, the typical interfacial width sets a minimum length 
scale in the system. For DPD, the natural scale is $r_c$, unless the 
parameters are chosen carefully.

\acknowledgements
The work of the FOM Institute is part of the research program of ``Stichting
Fundamenteel Onderzoek der Materie'' (FOM) and is supported by the
Netherlands Organization for Scientific Research (NWO). We acknowledge Pep
Espa\~{n}ol for sending us at the early stages of this work a preprint on a
similar model for treating conservative forces in DPD, P.B. Warren for
enlightening and encouraging discussions and J. Yeomans and S.I. Trofimov 
for helpful comments.



\begin{thebibliography}{99}
\bibitem{LB}  G.~R. McNamara and G. Zanetti, Phys. Rev. Lett. {\bf 61}, 2332
(1988).

\bibitem{LG}  U. Frisch, B. Hasslacher and Y. Pomeau, Phys. Rev. Lett. {\bf %
56}, 1505 (1986).

\bibitem{SPH}  J.~J. Monaghan, Annu. Rev. Astron. Astrophys. {\bf 30}, 543
(1992).

\bibitem{EMc}  D.~L. Ermack and J.~A. McCammon, J. Chem. Phys. {\bf 69},
1352 (1978).

\bibitem{BB}  J.~F. Brady and G. Bossis, Ann. Rev. Fluid. Mech.{\bf 20}, 111
(1988).

\bibitem{HK}  P.~J. Hoogerbrugge, and J.M.V. Koelman, Europh. Lett. {\bf 19}%
, 155 (1992).

\bibitem{Flekkoy}  E.~G. Flekk{\o}y and P.~V. Coveney, Phys. Rev. Lett. {\bf %
83}, 1775 (1999).

\bibitem{EW}  P. Espa\~{n}ol and P.~B. Warren, Europh. Lett. {\bf 30}, 191
(1995).

\bibitem{Colin}  C. Marsh, G. Bacxk, and M.~H. Ernst, Phys. Rev. E {\bf 56},
1676 (1997).

\bibitem{IF}  I. Pagonabarraga, M.~H.~J. Hagen and D. Frenkel, Europh. Lett. 
{\bf 42}, 377 (1998).

\bibitem{Masters}  A. Masters and P.~B. Warren, Europh. Lett. {\bf 48}, 1
(1999).

\bibitem{Serrano}  P. Espa\~{n}ol and M. Serrano, Phys. Rev. E {\bf 59},
6340 (1999).

\bibitem{pepL}  P. Espa\~{n}ol, Phys. Rev. E {\bf 57 }, 2930 (1998).

\bibitem{phasesep}  P.~V. Coveney and K.~E. Novik, Phys. Rev. E {\bf 54},
5134 (1996).

\bibitem{Evans}  R. Evans, in {\it Fundamentals of Inhomogeneous Fluids}, D.
Henderson ed., (Dekker, New York, 1992).

\bibitem{FS}  M.~W. Finnis, and J.~E. Sinclair, Philos. Mag. A, {\bf 50}, 45
(1984)

\bibitem{JY}  M.~R. Swift, E. Orlandini, W.~R. Osborn, and J.~M. Yeomans,
Phys. Rev. E {\bf 54}, 5041 (1996).

\bibitem{note}
If there is local structure, the last term in eq.(\ref{p:virial}) will have
an additional factor $(1+(1/d)[rw d\ln g(r)/dr]/[w])$. In this case, the
virial and thermodynamic pressures will differ. The true pressure is
the virial pressure, and the differences arise from correlations not
accounted for in the macroscopic free energy used to derive the thermodynamic expression for the pressure. However, if the local
structure varies smoothly, such differences can be disregarded.

\bibitem{H-H}  P.~C. Hohenberg and W.~P. Halperin, Rev. Mod. Physics {\bf 49}%
, 435 (1977).

\bibitem{Marini}  U. Marini Bettolo Marconi, and P. Tarazona, J. Chem. Phys. 
{\bf 110},8032 (1999).

\bibitem{Groot}  R.~D. Groot and P.~B. Warren, J. Chem. Phys. {\bf 107},
4423 (1998).

\bibitem{Godreche}  J. Langer, in {\it Solids far from Equilibrium}, C.
Godr\`{e}che ed., (Cambridge Univ. Press, Cambridge, 1991).

\bibitem{Sear}  R.~P. Sear, and W.~M. Gelbart, J. Chem. Phys. {\bf 110}, 458
(1999).

\bibitem{Sear1}  R.~P. Sear, S.~-W. Chung, G. Markovich, W.~M. Gelbart, and
J.~R. Heath, Phys. Rev. E {\bf 59}, R6255 (1999).

\bibitem{Irving}  J.~H. Irving and J.~G. Kirkwood, J. Chem. Phys. {\bf 18},
817 (1950).

\bibitem{Evans2}  R. Evans, Mol. Phys. {\bf 88}, 579 (1996).

\bibitem{Avalos}  J. Bonet Avalos, and A.~D. Mackie, Europh. Lett. {\bf 40},
141 (1997); P. Espa\~{n}ol, Europh. Lett. {\bf 40}, 631 (1997).
\end{thebibliography}
\end{document}